\documentclass[acmsmall,screen]{acmart}

\usepackage{tcolorbox}
\usepackage{float}
\usepackage{graphicx}
\usepackage{physics}
\usepackage{tabularx}

\AtBeginDocument{%
  }

\setcopyright{acmlicensed}
\copyrightyear{2024}
\acmYear{2024}
\acmJournal{TOSEM}
\acmVolume{00}
\acmNumber{00}
\acmArticle{00}
\acmMonth{00}

\usepackage{enumitem}

\newcommand{\arealabel}{ST}
\newlist{challenges}{enumerate}{1} 
\setlist[challenges,1]{
    label=\textbf{Ch-\arealabel-\arabic*.}, 
    ref=Ch-\arealabel-\arabic*, 
    leftmargin=*
}
\definecolor{LightGray}{rgb}{0.97,0.97,0.97}
\definecolor{DarkGray}{rgb}{0.25,0.25,0.25}

\begin{document}

\title{Quantum Software Engineering: Roadmap and Challenges Ahead}


\author{Juan M. Murillo}
\email{juanmamu@unex.es}
\orcid{0000-0003-4961-4030}
\author{Jose Garcia-Alonso}
\email{jgaralo@unex.es}
\orcid{0000-0002-6819-0299}
\author{Enrique Moguel}
\email{enrique@unex.es}
\orcid{0000-0002-4096-1282}
\affiliation{%
  \institution{Universidad de Extremadura}
  \country{Spain}
}

\author{Johanna Barzen}
\email{johanna.barzen@iaas.uni-stuttgart.de}
\orcid{0000-0001-8397-7973}
\author{Frank Leymann}
\email{frank.leymann@iaas.uni-stuttgart.de}
\orcid{0000-0002-9123-259X}
\affiliation{%
  \institution{University of Stuttgart. Institute of Architecture of Application Systems}
  \streetaddress{Institute of Architecture of Application Systems}
  \country{Germany}
}

\author{Shaukat Ali}
\email{shaukat@simula.no}
\orcid{0000-0002-9979-3519}
\affiliation{%
  \institution{Simula Research Laboratory}
  \country{Norway}
}

\author{Tao Yue}
\email{yuetao@buaa.edu.cn}
\orcid{0000-0003-3262-5577}
\affiliation{%
  \institution{Beihang University}
  \country{China}
}

\author{Paolo Arcaini}
\email{arcaini@nii.ac.jp}
\orcid{0000-0002-6253-4062}
\affiliation{%
  \institution{National Institute of Informatics}
  \country{Japan}
}

\author{Ricardo Pérez-Castillo}
\email{Ricardo.PdelCastillo@uclm.es}
\orcid{0000-0002-9271-3184}
\author{Ignacio García Rodríguez de Guzmán}
\email{Ignacio.GRodriguez@uclm.es}
\orcid{0000-0002-0038-0942}
\author{Mario Piattini}
\email{mario.piattini@uclm.es}
\orcid{0000-0002-7212-8279}
\affiliation{%
  \institution{University of Castilla-La Mancha}
  \country{Spain}
}

\author{Antonio Ruiz-Cortés}
\email{aruiz@us.es}
\orcid{0000-0001-9827-1834}
\affiliation{%
  \institution{I3US Institute, SCORE Lab, Universidad de Sevilla}
  \country{Spain}
}

\author{Antonio Brogi}
\email{antonio.brogi@unipi.it}
\orcid{0000-0003-2048-2468}
\affiliation{%
  \institution{University of Pisa}
  \country{Italy}
}

\author{Jianjun Zhao}
\email{zhao@ait.kyushu-u.ac.jp}
\orcid{0000-0001-8083-4352}
\affiliation{%
  \institution{Kyushu University}
  \country{Japan}
}

\author{Andriy Miranskyy}
\email{avm@torontomu.ca}
\orcid{0000-0002-7747-9043}
\affiliation{%
  \institution{Toronto Metropolitan University}
  \country{Canada}
}

\author{Manuel Wimmer}
\email{manuel.wimmer@jku.at}
\orcid{0000-0002-1124-7098}
\affiliation{%
  \institution{Johannes Kepler University Linz}
  \country{Austria}
}

\renewcommand{\shortauthors}{Murillo et al.}

\begin{abstract}
  
As quantum computers advance, the complexity of the software they can execute increases as well. To ensure this software is efficient, maintainable, reusable, and cost-effective —key qualities of any industry-grade software— mature software engineering practices must be applied throughout its design, development, and operation. However, the significant differences between classical and quantum software make it challenging to directly apply classical software engineering methods to quantum systems. This challenge has led to the emergence of Quantum Software Engineering as a distinct field within the broader software engineering landscape. In this work, a group of active researchers analyse in depth the current state of quantum software engineering research. From this analysis, the key areas of quantum software engineering are identified and explored in order to determine the most relevant open challenges that should be addressed in the next years. These challenges help identify necessary breakthroughs and future research directions for advancing Quantum Software Engineering.

\end{abstract}




\received{30 September 2024}
\received[revised]{31 October 2024}
\received[accepted]{30 November 2024}

\maketitle

%%%%%%%%%%%%%%%%%%%%%%%%%%%%%%%%%%%%%%%%%%%%%%%%%%
%%%%%%%%%%%%%%%%%%%%%%%%%%%%%%%%%%%%%%%%%%%%%%%%%%
\section{Introduction}
\label{sec:introduction}

Over the past two decades, quantum computing has rapidly transitioned from a theoretical concept to a burgeoning field of practical research and development. This transition has been driven by the advent of publicly accessible quantum computers, which have enabled researchers to experiment with algorithms that were previously confined to theoretical studies. Quantum algorithms have demonstrated the potential to solve certain problems in a reasonable timeframe that are beyond the capabilities of classical computers \cite{Serrano22}. These algorithms also offer significant operational advantages, such as the perfect training of Quantum Neural Networks using only a few highly entangled training data points \cite{Mandl}. The immense potential of quantum computing has captivated industries, leading to increased investment and the commercialization of quantum computers by manufacturers such as D-Wave, IBM, IonQ, Rigetti, and Quantinuum.

The current focus in quantum computing is on developing more stable Quantum Processing Units (QPUs) to move beyond the noisy intermediate-scale quantum (NISQ) era \cite{preskill2018quantum}. NISQ computers, despite their limitations \cite{Leymann_2020}, have already demonstrated their utility by solving problems through approximation methods, such as Variational Quantum Algorithms \cite{cerezo2021variational}. These advancements indicate that the era of industrialized quantum software is approaching, with the potential to revolutionize various fields by offering capabilities surpassing those of classical systems.

Nevertheless, historical lessons from software engineering reveal that the adoption of new technologies by the industry hinges on the ability to develop software in a repeatable, efficient, maintainable, reusable, and cost-effective manner \cite{ReusableComponents}. Currently, quantum software lacks the development procedures and standards prevalent in classical computing. Therefore, integrating sound engineering principles into quantum software is crucial to bridging this gap \cite{Piattini21-IT-pro,ali2022software,Zhao2020}.

Quantum Software Engineering (QSE) has been defined as \textit{``the use of sound engineering principles for the development, operation, and maintenance of quantum software\footnote{In this work, ``quantum software'' refers to software that runs on a quantum computer. In a broader sense, because all computing on quantum computers involves classical parts, software that utilizes quantum computers is naturally hybrid.}''} \cite{Zhao2020}. One of the main challenges in QSE is to translate the knowledge and practices from classical software engineering to the quantum domain while also developing new methodologies tailored to match the unique requirements of quantum computing \cite{stepney2004journeys}. A key aspect of this challenge is the anticipated hybrid model of quantum software, which will blend classical and quantum computing \cite{MiranskyyICSE19,weigold2021patterns}. This hybrid approach necessitates the development of technologies that enable seamless interaction between different computing environments. Leveraging Service-Oriented Computing (SOC) principles to create interoperable interfaces and adopting Service Engineering methodologies will be critical for the effective design and management of quantum services.

In addition, testing quantum systems presents unique challenges due to the peculiarities of quantum states \cite{TestingIntro}. Developing specific practices for quantum software testing is essential to ensure reliability and correctness. Additionally, requirements engineering for quantum software, although currently less emphasized due to the limited number of real-world applications, will inevitably diverge from classical approaches \cite{10313750,saraiva2021non}. Understanding and addressing these differences will be crucial as quantum applications will become more prevalent.

In summary, the path to widespread industrial adoption of quantum computing is fraught with challenges, but the ongoing advancements in QSE indicate a promising avenue for overcoming these hurdles. By focusing on the development of sound engineering principles and practices tailored to quantum software, the research community can facilitate the transition from the NISQ era to a future where quantum computing will be an integral part of the technological landscape \cite{AparicioMorales2024}. This journey will require concerted efforts in testing, requirements engineering, and the seamless integration of classical and quantum systems, ultimately leading to the realization of practical and scalable quantum software solutions.

This paper aims to provide a comprehensive overview of the primary challenges arising in different areas of Quantum Software Engineering. It highlights recent research developments in QSE and discusses the future challenges that researchers are likely to face in the coming years. By addressing these challenges, the field of QSE can advance significantly, paving the way for the development of robust, efficient, and maintainable quantum software. The future of quantum computing holds immense promise, and the progress in QSE will be instrumental in realizing the full potential of this revolutionary technology.

To this end, this paper is structured as follows. Section \ref{sec:fundamentals} presents some key concepts of quantum computing. Section \ref{sec:research_interests} provides a wide exploration of the research interest in the field of QSE. Section \ref{sec:research_areas} shows a detailed review of the different research areas of QSE. Section \ref{sec:keyAreas} discusses the different key areas of software engineering and how they can influence quantum computing. Finally, Section \ref{sec:conclusion} presents some implications and conclusions drawn from this research.

%%%%%%%%%%%%%%%%%%%%%%%%%%%%%%%%%%%%%%%%%%%%%%%%%%
%%%%%%%%%%%%%%%%%%%%%%%%%%%%%%%%%%%%%%%%%%%%%%%%%%
\section{Fundamentals of Quantum Computing}
\label{sec:fundamentals}

For those readers unfamiliar with quantum computing, this section introduces some basic concepts that will be useful for understanding the remaining sections. Quantum computing takes advantage of distinctive properties and processes rooted in the fundamental principles of quantum physics, setting it apart as a completely different paradigm from classical computing.

At the heart of quantum computing, there are key principles derived from quantum mechanics, including the use of quantum bits, or qubits, along with phenomena like superposition and entanglement. These quantum properties, along with many others, provide novel opportunities to enhance computational capabilities in ways that classical systems cannot achieve.

%%%%%%%%%%%%%%%%%%%%%%%%%%%%%%%%%%%%%%%%%%%%%%%%%%
\subsection{Qubit}

In classical computing, the basic unit of information is the bit, which can take on the values of either 0 or 1. In quantum computing, however, we have the qubit (quantum bit), which serves as the fundamental unit of information.

A qubit can exist in a state of 0, 1, or a superposition of both simultaneously. This superposition enables quantum computers to perform complex computations more efficiently than classical computers. Mathematically, a qubit is often represented using Dirac notation, introduced by physicist Paul Dirac \cite{dirac}. In this formalism, a qubit's state is expressed as a vector with two complex components, written as:

\begin{equation}
    |\psi\rangle = \alpha|0\rangle + \beta|1\rangle
    \label{equation:dirac}
\end{equation}

In this equation, $|\psi\rangle$ is the state of the qubit, and $\alpha$ and $\beta$ are complex numbers representing the probability amplitude of finding the qubit in the $|0\rangle$ and $|1\rangle$ basis states, respectively.  These states are the computational basis of the qubit and are analogous to the classical 0 and 1.

%In addition, it must be satisfied that the magnitude squared of the probability amplitudes $\alpha$ and $\beta$ sum to 1, 
In addition, the squared magnitude of the amplitudes $\alpha$ and $\beta$ must sum to 1, which means that the probability of finding the qubit in one of the two states is 1. This can be represented as follows:

\begin{equation}
    |\alpha|^2 + |\beta|^2 = 1
    \label{equation:magnitudes}
\end{equation}

%In addition to the mathematical representation of the qubit, it is common to use the Bloch Sphere for its graphical representation, as can be seen in Figure \ref{fig:blochSphere}. In this spherical representation, each point on the sphere corresponds to a single quantum state of the qubit. The upper pole represents the $|0\rangle$ state, while the lower pole represents the $|1\rangle$ state, and the intermediate points on the surface of the sphere represent the superposition of the two base states of the qubit.

%The points on the sphere can be described mathematically using Equation \ref{equation:dirac}, where the coordinates are related to the amplitudes of the qubit.

%\begin{figure}[!ht]
%    \centering
%    \includegraphics[width=0.3\textwidth]{images/BlochSphere.png}
%    \caption{Bloch Sphere}
%    \label{fig:blochSphere}
%\end{figure} 

%%%%%%%%%%%%%%%%%%%%%%%%%%%%%%%%%%%%%%%%%%%%%%%%%%
\subsection{Superposition}

Superposition enables quantum computers to perform parallel computations in a single step, as a qubit in a superposition state can execute operations on all its possible states simultaneously \cite{Bouwmeester2000}. For instance, if two qubits are in superposition, they can collectively represent four different states (00, 01, 10, and 11) simultaneously, whereas two classical bits could only represent one of these states at a time. As the number of qubits increases, the ability to represent multiple states grows exponentially, following the formula $States=2^{n}$, where $n$ is the number of qubits.

However, when a measurement is performed on a qubit in superposition, that is, when its value is observed, the superposition collapses. The qubit assumes one of its classical states (either 0 or 1), with the outcome determined by the probability amplitudes of each state. This phenomenon is a fundamental aspect of quantum computing: until a measurement is made, the qubit exists in a superposition of states, and only upon observation does it take on a definite value \cite{LaPierre2021}.

%%%%%%%%%%%%%%%%%%%%%%%%%%%%%%%%%%%%%%%%%%%%%%%%%%
\subsection{Entanglement}

Quantum entanglement is a fundamental concept in quantum physics, crucial to the functioning of quantum computing. It describes a unique correlation between entangled particles, such as qubits, that are so deeply interconnected that the state of one qubit directly influences the state of the other, regardless of the distance between them. This phenomenon challenges classical notions of independence between separate entities, where the state of one object does not affect another unless they are in direct interaction \cite{Horodecki2009}.

In this context, when two qubits become entangled, the state of one qubit is intrinsically linked to the state of its partner. Altering the state of one qubit immediately changes the state of the other, even if they are separated by vast distances. This non-locality enables the formation of entangled qubit pairs, which are critical for many quantum computational processes.

Entanglement also introduces the concept of non-local interactions, where measuring one qubit in an entangled pair instantly reveals information about the other qubit, regardless of their spatial separation. This property enhances the speed and efficiency of certain quantum algorithms to the point that, without entanglement, an exponential increase in speed cannot be achieved \cite{Jozsa2003}.

%, as the qubits share information instantaneously across distances. However, similar to superposition, the exact state of an entangled qubit is only determined upon measurement. Until observed, the qubits exist in a correlated state, demonstrating the complex and somewhat counterintuitive nature of quantum entanglement. This phenomenon is fundamental to many quantum computing protocols and is also the basis for emerging quantum communication technologies. And more importantly for quantum computing , without entanglement, an exponential increase in speed cannot be achieved \cite{Jozsa2003}. 

%One of the practical applications of entanglement appears in quantum cryptography \cite{Mavroeidis2018}. In this domain, entanglement ensures secure communication, as any attempt to eavesdrop or interfere with the quantum state would disturb the entanglement, revealing the presence of an intrusion. This property provides a powerful method for generating secure quantum encryption keys, forming the basis for future quantum encryption systems.

%%%%%%%%%%%%%%%%%%%%%%%%%%%%%%%%%%%%%%%%%%%%%%%%%%
\subsection{Quantum state vector and quantum register}

A quantum state vector (also simplified as a quantum state or state vector) is a mathematical representation that encapsulates the information about a quantum system. It contains all the information needed to predict the outcomes of measurements on the system.

In this context, a quantum state refers to the state of a quantum system as it evolves during the execution of a quantum algorithm. It describes the configuration of qubits at a particular point during the execution of the quantum program. To demonstrate these concepts, consider a quantum system consisting of two qubits. The state vector $\mathbf{|\psi\rangle}$ for this system can be expressed as:

\[
\mathbf{|\psi\rangle} = \begin{pmatrix}
c_{00} \\
c_{01} \\
c_{10} \\
c_{11}
\end{pmatrix} = \begin{pmatrix}
c_{00} & c_{01} & c_{10} & c_{11}
\end{pmatrix}^T
\]

In this case, each $c_{ij}$ represents the amplitude corresponding to the basis state $\vert ij \rangle$. The basis states $\vert 00 \rangle$, $\vert 01 \rangle$, $\vert 10 \rangle$, and $\vert 11 \rangle$ form a four-state orthonormal basis. For example, if the system is in the state $|\psi\rangle = (1  0  0  0)^T$,  it means the amplitude for the state $\vert 00 \rangle$ is $1$  and $0$ for the remaining states $\vert 01 \rangle$, $\vert 10 \rangle$ and $\vert 11 \rangle$. Consequently, the probability of measuring each state on a computational basis is determined by the squared modulus of their corresponding amplitudes. In this case $|1|^2 = 1$ for state $\vert 00 \rangle$ and $|0|^2 = 0$ for the remaining states. Thus, the system has a 100\% chance of being measured in the state $\vert 00 \rangle$ and no chance of being measured in the states $\vert 01 \rangle$, $\vert 10 \rangle$, or $\vert 11 \rangle$.

In more complex quantum systems, with more qubits involved, sets of qubits can be conceptually grouped into quantum registers. A quantum register is the quantum analogue of a classical register, playing a key role in storing and processing quantum information during a computation. In these complex systems, each quantum register can be represented by its own quantum state vector.

%%%%%%%%%%%%%%%%%%%%%%%%%%%%%%%%%%%%%%%%%%%%%%%%%%
\subsection{Circuit-based quantum computing and Quantum annealing}

Currently, two different kinds of quantum computers are available for researchers interested in building quantum software: quantum annealers and circuit-based quantum computers. Quantum annealers specialize in solving optimization problems by finding the lowest energy state of a system. Circuit-based quantum computers are general-purpose quantum computers capable of running various algorithms. The following fundamental concepts are related to circuit-based quantum computers since, being general-purpose computers, they have gathered more interest from the quantum software engineering community. Nevertheless, quantum annealing works will also be analyzed and discussed during the rest of the paper. 

%%%%%%%%%%%%%%%%%%%%%%%%%%%%%%%%%%%%%%%%%%%%%%%%%%
\subsection{Quantum Gate}

Quantum gates are core components of quantum software, similar to classical logic gates. Unlike classical gates that work with bits, quantum gates manipulate qubits. They alter the quantum states of qubits by adjusting their probabilities of being measured in specific states or by creating entanglement between qubits.  Mathematically, each gate is represented by a $2^{n} \times 2^{n}$ unitary matrix where $n$ is the number of qubits affected by the gate. The action of a gate on a specific quantum system is determined by multiplying the state vector by the matrix representation of the gate.

\begin{figure}[!ht]
    \centering
    \includegraphics[width=0.5\textwidth]{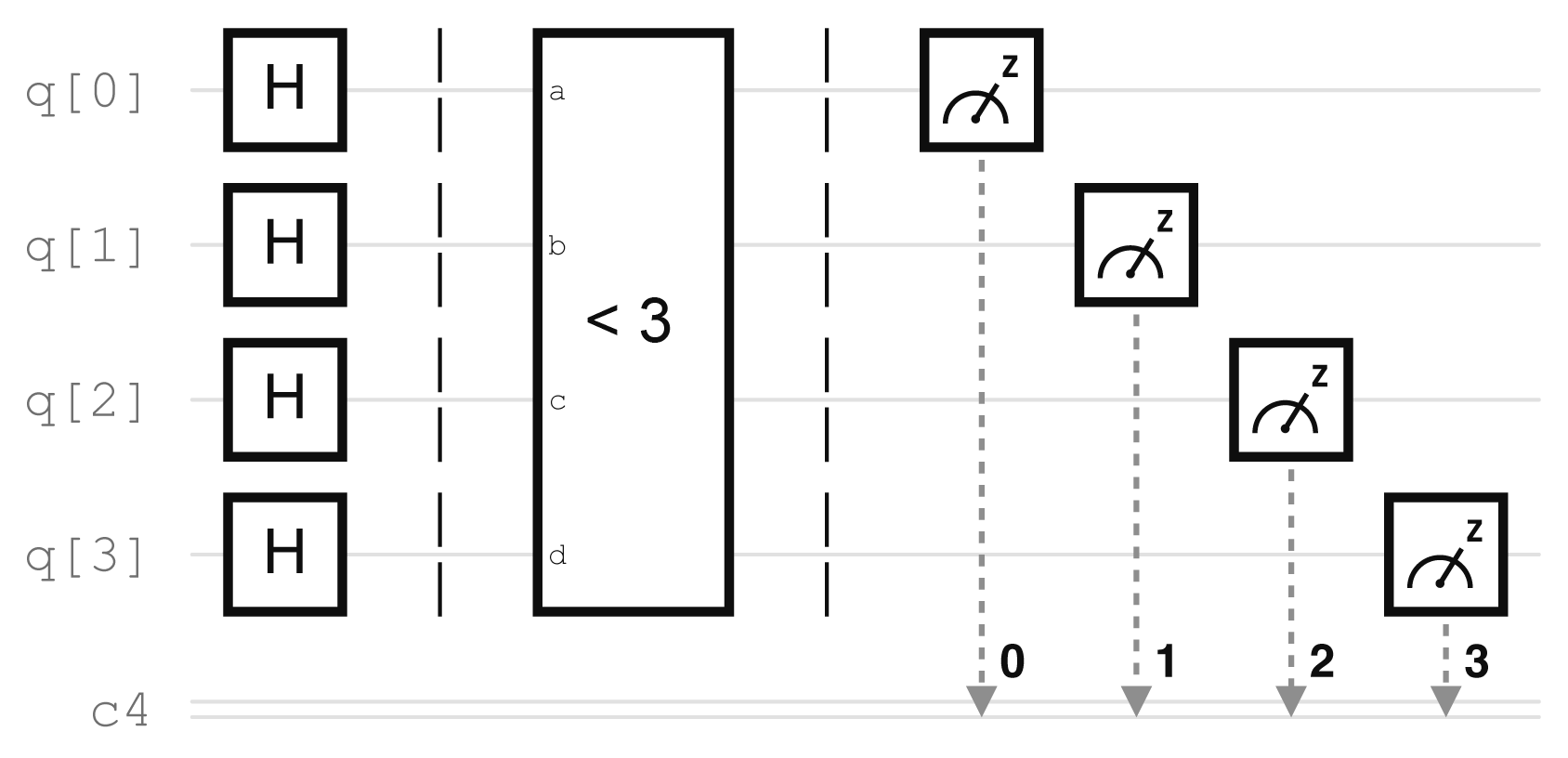}
    \caption{Example of a gate-based quantum circuit}
    \label{fig:ciruit}
\end{figure}

Notable examples include the Hadamard gate for creating superposition (gate H in Figure \ref{fig:ciruit}), the Pauli-X gate (which flips qubit states), and the CNOT gate (for controlled operations creating entanglement between qubits). These gates are essential for executing quantum algorithms, enabling complex manipulations and transformations within a quantum system.

%%%%%%%%%%%%%%%%%%%%%%%%%%%%%%%%%%%%%%%%%%%%%%%%%%
\subsection{Quantum Circuit}

A quantum circuit is a sequence of quantum gates arranged to manipulate qubits in a specific order, enabling the execution of quantum algorithms. It consists of multiple quantum gates, measurement operations, and potentially classical logic, all organized to perform a desired computation. Figure \ref{fig:ciruit} illustrates an example of a gate-based quantum circuit. More specifically, the circuit shown in Figure \ref{fig:ciruit} is a circuit that gives all numbers less than three for four qubits in a superposition state. Such a circuit yields a result (0000, 0001, and 0010).

Quantum circuits are graphically represented with qubits as horizontal lines, gates as symbols that indicate operations applied to these qubits, and modules encapsulating a set of gates like the oracle in the figure (oracle with the operation less than \textit{n}). By altering the quantum states through operations like superposition and entanglement, a quantum circuit can perform complex computations that would be infeasible for classical circuits, making quantum circuits fundamental for implementing quantum algorithms.

To perform the computations they are designed for, quantum circuits are run (executed) on quantum machines. These quantum machines can be real quantum hardware or quantum simulators, and software systems can perform quantum computations using classical hardware. Quantum simulators can be ideal (meaning that quantum computations are executed perfectly) or include noise models to replicate the behaviour of real quantum hardware. In any case, simulators are limited in terms of the number of qubits that can be simulated in classical hardware.

%%%%%%%%%%%%%%%%%%%%%%%%%%%%%%%%%%%%%%%%%%%%%%%%%%
\subsection{Quantum Algorithm}

A quantum algorithm is a structured sequence of operations tailored to run on a quantum computer, leveraging quantum properties such as superposition or entanglement to solve computational problems that are infeasible or highly inefficient for classical computers. By manipulating qubits through quantum gates, these algorithms explore multiple solutions simultaneously and perform complex operations in parallel, enabling them to tackle problems with significantly reduced computational complexity.

Quantum algorithms exploit the probabilistic nature of quantum mechanics to achieve speedups over classical methods for certain types of tasks. For example, Shor’s algorithm \cite{Shor1999} provides an exponential speedup for factoring large integers, reducing the time complexity from sub-exponential (classical) to polynomial (quantum). It does this by converting the problem into a period-finding task and using the Quantum Fourier Transform (QFT) \cite{Weinstein1889} to efficiently determine the periodicity of a function, a critical step in factorization. This capability threatens encryption systems like RSA, which rely on the classical difficulty of factoring large numbers, showcasing quantum computing’s potential to disrupt current cryptographic standards.
%Shor’s Algorithm \cite{Shor1999} can factorize large integers in polynomial time, breaking the security of widely used cryptographic systems based on the difficulty of factorization, for which classical algorithms require exponential time. 

Similarly, Grover’s algorithm \cite{Grover1996} achieves a quadratic speedup for unstructured search problems. Classical search methods require \textit{O(N)} time to find an item in an unsorted database, but Grover’s algorithm reduces this to \textit{$O(\sqrt{N})$} using amplitude amplification. By encoding all possible solutions into quantum superpositions and iteratively amplifying the correct solution’s amplitude, the algorithm identifies the solution in far fewer steps. Although the speedup is not exponential, it provides a meaningful advantage for large-scale search problems, highlighting quantum computing’s utility in domains where classical approaches face scalability challenges.

Together, these algorithms illustrate the transformative potential of quantum computing in solving problems more efficiently than classical methods.

%%%%%%%%%%%%%%%%%%%%%%%%%%%%%%%%%%%%%%%%%%%%%%%%%%
\subsection{Quantum Oracle}
\label{quantumOracle}

An oracle is a piece of circuit performing a function that, applied to a quantum state, produces as output the same quantum states with some transformations. This output will be used as input for another circuit that implements some algorithm.  Thus, oracles can be thought of as a black box performing a function that is used as an input by another algorithm constituting a pattern for quantum algorithms \cite{Leymann-QuantumAlgorithms}. Oracles appear in many well-known quantum algorithms like Grover \cite{Grover_1998},  Deutsch-Jozsa \cite{Deutsch1992RapidSO}, Simon \cite{Simon_1997} or Bernstein-Vazirani \cite{Bernstein_Vazirani}.

%%%%%%%%%%%%%%%%%%%%%%%%%%%%%%%%%%%%%%%%%%%%%%%%%%
\subsection{Quantum Noise}
\label{quantumNoise}

Noise in quantum computers and simulators refers to any unwanted interaction or imperfection that disrupts the delicate quantum states used for computation. Quantum systems rely on phenomena like superposition and entanglement, which are inherently fragile and highly sensitive to external influences. Noise manifests as errors in quantum computations, leading to incorrect results or the loss of coherence in quantum states. The sources of noise in quantum systems can be broadly categorized as follows:

\begin{itemize}
    \item Environmental Noise. Quantum systems are highly susceptible to interactions with their surroundings, such as electromagnetic radiation, thermal fluctuations, and vibrations. These external disturbances can cause decoherence, a process where qubits lose their quantum properties and revert to classical states.

    \item Gate Errors. Quantum gates are not perfect, and inaccuracies during their execution introduce gate errors. These arise due to hardware imperfections, such as imprecise control over qubits, timing issues, or limitations in the precision of control signals.

    \item Measurement Errors. Reading the state of a qubit (measurement) is another source of noise. Due to the probabilistic nature of quantum mechanics and imperfections in the measurement apparatus, the observed state may not accurately reflect the actual state of the qubit.

    \item Cross-Talk. In multi-qubit systems, operations on one qubit can inadvertently influence neighboring qubits due to unintended coupling or interference, leading to errors.

    \item Qubit Relaxation and Dephasing. Qubits have finite coherence times, meaning they can only maintain their quantum states for a limited duration. Relaxation occurs when a qubit spontaneously transitions to its ground state, while dephasing refers to the loss of relative phase information between superposed states.
\end{itemize}

On the other hand, in quantum simulators, noise can be artificially introduced to mimic the imperfections of real quantum hardware. This is essential for testing and developing error mitigation and correction strategies, which aim to counteract the effects of noise in practical quantum computing.

\section{Research interest in Quantum Software Engineering}
\label{sec:research_interests}

% \note{REVIEWER COMMENT: In the full paper, I also suggest providing a brief overview of how quantum software differs from traditional software in general, so that the paper is more accessible to readers who are not already working in this area.}

QSE has evolved into a distinct discipline within the broader field of software engineering. First mentioned in 2002, QSE was identified as one of the grand challenges in computer science research. Clark \textit{et al.} described it as ``\textit{the development of a full discipline of Quantum Software Engineering, ready to exploit the full potential of commercial quantum computer hardware, once it arrives, projected to be around 2020}'' \cite{Clark2002}.

Although the early motivation for developing QSE was driven by the anticipated advancements in quantum hardware, it took several years before quantum hardware and simulators became available to the public. A significant milestone was reached in 2016 when IBM released its first gate-based quantum computer along with an online simulator \cite{IBM2016Newsroom}. This breakthrough was soon followed by other vendors, catalyzing the growth and development of QSE.

Since the early developments in Quantum Software Engineering, several conference venues have emerged that significantly contribute to the growth of the field. While not all conferences focus solely on QSE, many include papers on the topic, reflecting the growing interest and research in this area.

One such notable event is the \textit{International Workshop on Quantum Software Engineering (Q-SE)}. This workshop, which celebrated its fifth edition this year, is collocated with the \textit{ACM/IEEE International Conference on Software Engineering (ICSE)}. Another important conference is the \textit{IEEE International Conference on Quantum Software (QSW)}. Now in its third edition, this conference is organized under the umbrella of the \textit{IEEE World Congress on Services}, providing a platform for discussing advancements in quantum software. Additionally, the \textit{International Workshop on Quantum Software Engineering and Technology (Q-SET)} is a significant event within the \textit{IEEE Quantum Week}. This workshop focuses on advancing the field of QSE by bringing together researchers and practitioners to share their latest findings. The \textit{IEEE International Conference on Quantum Computing and Engineering (QCE)}, also part of \textit{IEEE Quantum Week}, is another prominent venue that addresses various aspects of quantum computing, including software engineering.

These conferences and workshops provide essential platforms for researchers and practitioners to share advancements, discuss challenges, and foster collaboration in the field of Quantum Software Engineering. Through these gatherings, the community can collectively push the boundaries of what is possible with quantum software, paving the way for future innovations.

On the other hand, the interest in QSE extends deeply into the realm of academic journals, where the importance of this emerging discipline is highlighted through dedicated issues and sections. Among these, the \textit{IEEE Transactions on Quantum Engineering} and \textit{ACM Transactions on Quantum Computing} are prominent journals that cover a broad spectrum of topics, including QSE. Furthermore, Elsevier \textit{Journal of Systems and Software} and \textit{Information and Software Technology} have recognized the growing significance of QSE by introducing special issues specifically dedicated to this area. Additionally, \textit{ACM Transactions on Software Engineering and Methodology} features a Continuous Special Section on Quantum Software Engineering. 

The inclusion of QSE in these prestigious journals underscores the field's growing importance and the need for rigorous academic inquiry. These publications serve as critical venues for researchers to share their findings, discuss emerging trends, and collaborate on addressing the complex challenges associated with developing robust and effective quantum software. Through these journals and special issues, the body of knowledge in Quantum Software Engineering continues to expand, driving the field forward.

\begin{figure}[!ht]
    \centering
    \includegraphics [width=0.8\columnwidth]{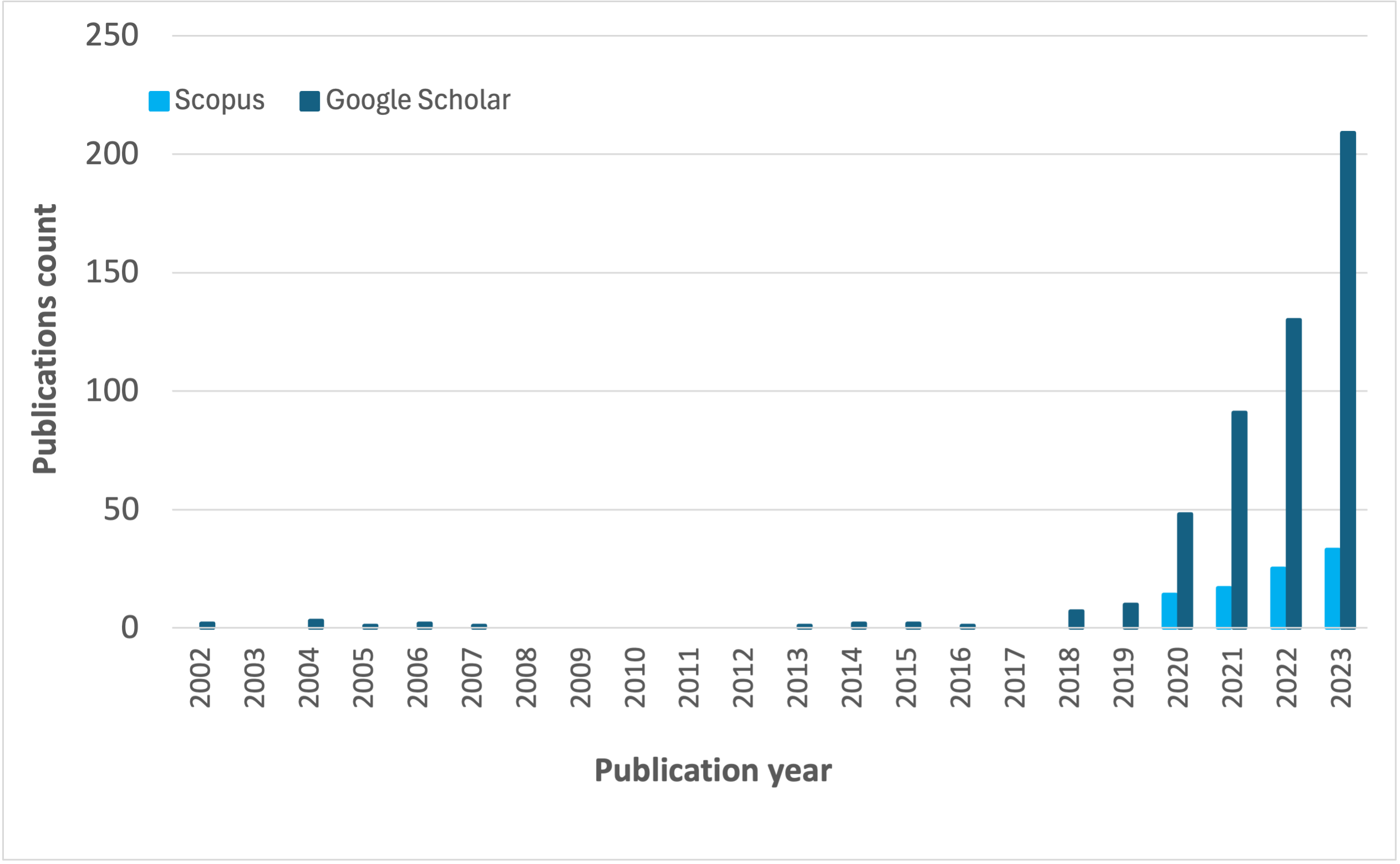}
    \caption{Papers on QSE over time}
    \label{fig:numberPublications}
\end{figure}

The rising interest in QSE can be indirectly measured by analyzing the number of publications containing the term ``\textit{Quantum Software Engineering}'', taking into account as inclusion criteria all types of papers, proceedings, books, technical reports, etc. Figure \ref{fig:numberPublications} illustrates the annual number of such publications indexed in Scopus and Google Scholar\footnote{For replication purposes, the specific publications considered are listed in \url{https://doi.org/10.5281/zenodo.13839576}.}. Some initial publications appeared between 2002 and 2007, highlighting the early recognition of the need for QSE in the emerging quantum computing domain. A resurgence of publications began around 2013, primarily driven by the needs identified by quantum computing scientists during the development of quantum software. The availability of quantum simulators and platforms around 2020 led to a significant increase in QSE publications, with more than 200 publications recorded in 2023. % And more than 100 publications accounted for until 31 July 2024.

Focusing on the papers found in the Scopus database, probably more mature than its counterpart in Google Scholar, of the 102 papers identified up to 31 July 2024, 70 (68.6\% of the total) were conference publications, 26 (25.5\%) were journal articles and 6 (5.9\%) were book chapters. The data clearly indicates that conferences are the main venue for the publication of research papers in QSE, although it should be emphasized that of the 26 journal publications, 10 works (more than a third of the total) have appeared in the first half of 2024.

\begin{figure}[!ht]
    \centering
    \includegraphics [width=0.9\columnwidth]{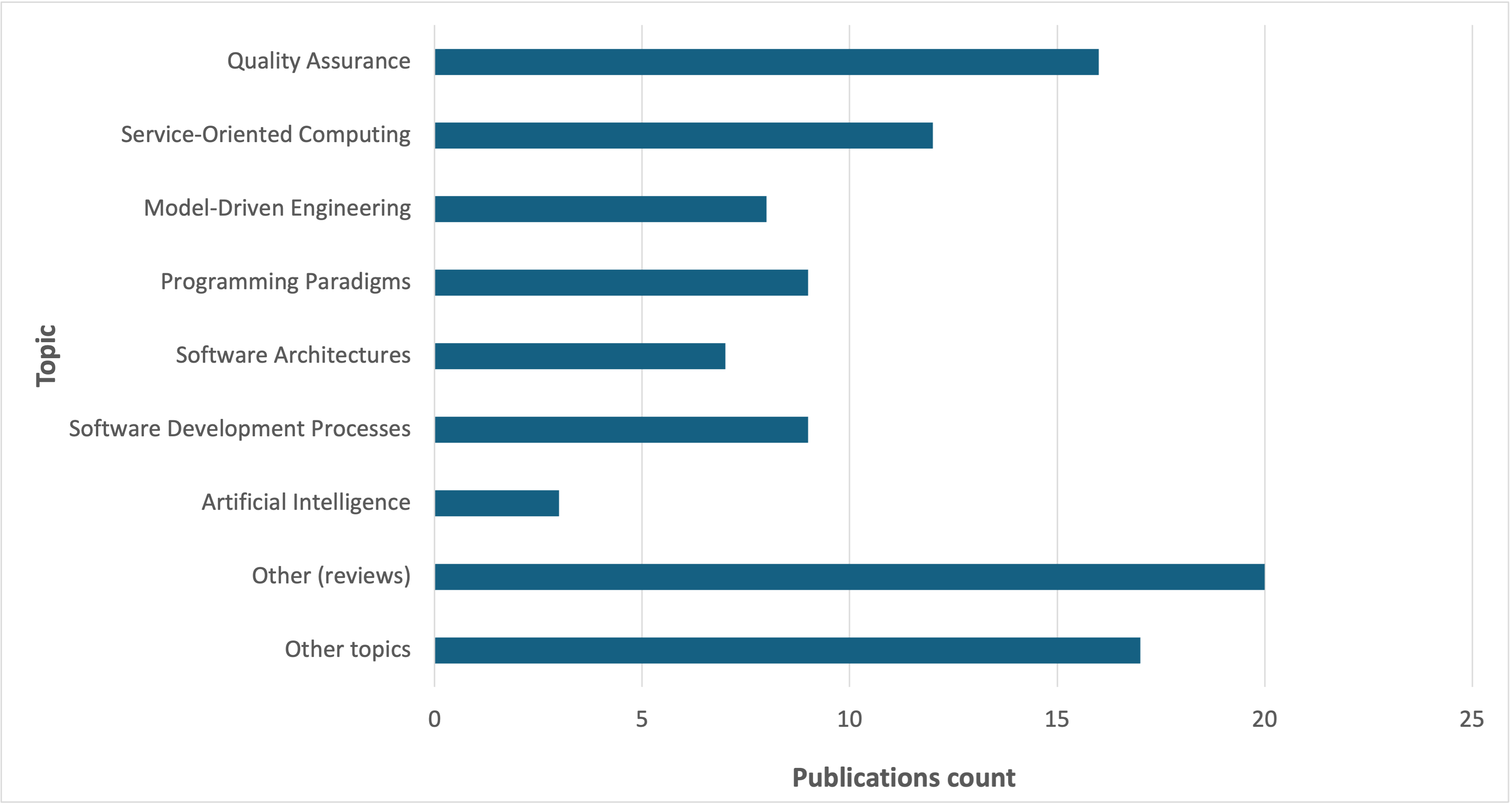}
    \caption{Number of publications by topic}
    \label{fig:publicationsByTopic}
\end{figure}

Once all these works have been analyzed and classified by the different areas of knowledge in the field of QSE, we obtain the results shown in Figure \ref{fig:publicationsByTopic}. It is noteworthy that the topics with the highest number of publications are ``\textit{Quality Assurance}'' with 16 publications and ``\textit{Service-Oriented Computing}'' with 12 publications. We also find the topics ``\textit{Programming Paradigms}'' and ``\textit{Software Development Processes}'' both with 9 works, ``\textit{Model-Driven Engineering}'' with 8 publications, ``\textit{Software Architectures}'' with 7 articles, and ``\textit{Artificial Intelligence}'' with 3 publications. It is also important to note that from this analysis we have found 20 review publications (surveys, systematic literature reviews, systematic mapping studies, and other types of reviews). In addition, we identified 17 papers on topics of less relevance to QSE such as communications, cybersecurity, or requirements engineering.

%%%%%%%%%%%%%%%%%%%%%%%%%%%%%%%%%%%%%%%%%%%%%%%%%%
%%%%%%%%%%%%%%%%%%%%%%%%%%%%%%%%%%%%%%%%%%%%%%%%%%
\section{Quantum Software Engineering Active Research Areas}
\label{sec:research_areas}

% \note{REVIEWER COMMENT: In general, the full paper should expand the discussion of each research area. In particular, in the full paper, it would be good to give clear and explicit examples to illustrate the problems in each area. For example, it is not explicitly clear how traditional approaches to Service-Oriented Computing fail to apply to quantum software. Use of a concrete example application may help illustrate why this challenge is important. Similarly, in the Architecture section, examples of design patterns proposed for quantum/hybrid software would be very illuminating}

This section examines some of the research conducted in different areas of QSE. We do not aim to provide a systematic analysis. Instead, we strive to gather QSE works and opinions from active researchers in the field regarding the most relevant challenges they anticipate in QSE for the upcoming years.

%%%%%%%%%%%%%%%%%%%%%%%%%%%%%%%%%%%%%%%%%%%%%%%%%%
\subsection{Quality Assurance: Testing and Debugging}
\label{sec:testing}

Quantum software testing is about assessing the correctness of quantum software in delivering their intended functionalities, while debugging is about observing failures in quantum software that need to be diagnosed and fixed, especially when facing challenges such as quantum noise, limited quantum hardware and simulator resources, and limited observability \cite{ali2022software}.

Several groups have made notable contributions to test quantum software in different ways \cite{MiranskyyICSE19,MiranskyyICSE2020,miranskyy2021testing,long2024testing,ye2023quratest}. Various coverage criteria for assessing the quality of test cases have been proposed, including covering inputs and outputs of quantum programs \cite{ali2021assessing,wang2021quito}, coverage criteria based on equivalence class partitioning \cite{long2024testing}, and multi-granularity coverage criteria for Quantum Neural Networks (QNNs) \cite{shao2024coverage}. In addition, several advanced techniques have been applied to test quantum programs including search-based testing \cite{wang2021generating,qusbtICSE22}, combinatorial testing \cite{Combinatorial,QuCat}, metamorphic testing \cite{Rui_metamorphic}, fuzzing \cite{wang2018quanfuzz}, property-based testing \cite{honarvar2020property}, concolic testing \cite{xia2024concolic}, and also techniques based on the reversibility property of quantum circuits \cite{GarciaDeLaBarrera2023}. Moreover, techniques for testing quantum computing platforms have also emerged, such as those based on metamorphic testing and differential testing \cite{morphq,Qdiff}. Various assertion types have been suggested to check program output correctness, such as statistical assertions \cite{StatisticalAssertions}, projection-based assertions \cite{ProjectAssertions}, and dynamic assertions \cite{li2014debugging, zhou2019quantum, DynamicAsseertions}. Moreover, as quantum computer noise makes it difficult to distinguish actual bugs from program failures caused by noise, efforts have recently been made to mitigate noise in quantum computing to enhance testing reliability (e.g., see \cite{AsmarTSE}). A testing framework \cite{long2024testing} has also been developed to support both unit and integration testing of quantum programs. Relations between quantum programs are also leveraged to address the test oracle\footnote{This should not be confused with the definition of oracle seen in Section \ref{quantumOracle} above.} problem in quantum software testing \cite{long2024equivalence}.

Mutation testing is an efficient approach to assess the quality of a test suite. Different works have proposed mutation operators for quantum programs, like Muskit \cite{muskitASE21tool} and QmutPy \cite{QmutPy,QMutPy2,QMutPy3}. Moreover, an empirical evaluation was conducted on over 700,000 mutants generated from more than 350 real quantum circuits \cite{usandizaga2023quantum}, and the generation approach MutTG was proposed for creating minimal-sized test suites capable of killing a set of mutants \cite{WangGECCO2022}.

An empirical analysis of software repositories containing quantum programs is underway, which leads to the identification of bug patterns (some of which are quantum-specific) affecting quantum software \cite{zhao2021identifying,paltenghi2022bugs,zhao2023bugs4q,zhao2023empirical,campos2021qbugs}. These patterns can be used by newly-created static analyzers for quantum code, such as \cite{paltenghiFSE2024lintq,kher2023automatic,nayak2023qpac,zhao2023qchecker,xia2023static}, to detect defects early in the development process. Additionally, they can be used to create benchmarks for testing and debugging \cite{zhao2023bugs4q}. Quantum software bug trackers reveal flaky tests (which is intuitive given quantum mechanics' probabilistic nature), but the root causes of flakiness differ from classical software \cite{zhang2023identifying,zhang2024automated,sivaloganathan2024automating}. Transformer-based models (ranging from CodeBERT~\cite{feng2020codebert} to large language models) are being explored to classify bug reports and detect defects in quantum bug repositories by leveraging information from bug reports, change requests, and source code \cite{sivaloganathan2024automating}. Quantum-specific technical debt \cite{openja2022technical} and code smells \cite{chen2023smelly} are being analyzed and detected, and software repositories are used to validate the findings empirically. Lastly, software repositories can be used to identify a list of developers to be interviewed, e.g., to assess their background \cite{shaydulin2020making} or elicit code smells \cite{chen2023smelly}. Formal methods, such as the formalization of Quantum Intermediate Representation (QIR), also contribute to validating program safety and detecting unsafe operations \cite{luo2023enhancing,luo2025formalization}. Moreover, an investigation into automatic quantum program repair has begun, demonstrating the effectiveness of large language models, specifically with OpenAI's GPT-3, in repairing some quantum programs \cite{guo2024repairing}. Li \textit{et al.}~\cite{li2024automatic} proposed a framework for repairing quantum programs using unitary operations. By systematically generating and validating patches, the framework addresses defects in quantum programs while ensuring correctness through algebraic principles. Additionally, Yu and Zhao~\cite{yu2025qpdg} introduced the Quantum Program Dependence Graph (QPDG), which effectively models both quantum-specific and classical dependencies in quantum programs. This framework facilitates slicing and dependency analysis, aiding in the debugging, testing, and understanding of quantum programs.

Debugging involves analyzing programs to identify and correct errors through causal analysis between bugs and detected errors, crucial for both classical and quantum programming \cite{bentley1985programming,chong2017programming}. Miranskyy \textit{et al.} \cite{MiranskyyICSE2020,miranskyy2021testing} examined quantum program debugging tactics, exploring the adaptation of classical strategies—brute force, backtracking, and cause elimination—to quantum contexts. Metwalli and Meter \cite{metwalli2024testing} proposed a debugging framework for quantum circuits, concentrating on amplitude permutation, phase modulation, and amplitude redistribution circuit blocks, addressing the need for specialized approaches for each quantum circuit type, contributing towards robust quantum computing systems. Sato and Katsube \cite{sato2023locating} introduced four considerations for locating bugs in quantum programs and an efficient bug-locating method integrating cost-based binary search, early determination, finalization, and retrospection, validated by experimental results. Recently, delta debugging has been explored in the context of regression testing of quantum programs \cite{DeltaDebugging} with a specific focus on property-based testing. 

In quantum program debugging, implementing quantum assertions often involves duplicating quantum variables and repeated measurements. Huang \textit{et al.} \cite{StatisticalAssertions} used hypothesis testing to partially reconstruct quantum register information from measurements for assertions, noting a lack of runtime assertion support. To overcome this, Liu \textit{et al.} \cite{DynamicAsseertions} proposed using additional qubits to capture assertion-target qubits' information. Furthermore, Li \textit{et al.} \cite{ProjectAssertions} introduced Proq, a projection-based runtime assertion tool, utilizing projection measurements to enable assertions without destroying quantum variables. Additionally, Jin and Zhao proposed ScaffML \cite{jin2023scaffml}, a behavioral interface specification language for the quantum programming language Scaffold \cite{abhari2012scaffold}, which specifies pre- and post-conditions for Scaffold modules to support the debugging and testing of Scaffold programs. The researchers have also realized the need to compress program specifications often used to support automated testing. For instance, Oldfield \textit{et al.} \cite{oldfield2024fasterbetterquantumsoftware} proposed an algorithm to compress program specifications to improve testing efficiency.  

Given the noise on current quantum computers, testing quantum programs on real quantum computers or their corresponding noisy simulators poses additional challenges. For instance, a tester may be unable to reliably determine whether a test case failed a program or was due to noise in quantum computations. To this end, a set of approaches were recently proposed \cite{AsmarFSE,AsmarTSE,pauliStringsASE2024,MuqeetSSBSE2024}. Muqeet \textit{et al.} \cite{AsmarFSE,AsmarTSE} proposed two machine learning-based approaches to learn noise patterns as machine learning models for a given quantum computer or noisy simulator. Such machine learning models are then used to filter out noise from the outputs of quantum programs, thereby improving testing reliability. Moreover, to enable the use of existing error mitigation techniques implemented on IBM's quantum computers, an approach has been proposed by Muqeet \textit{et al.} \cite{pauliStringsASE2024}. Naturally, the approach can be used to support testing on quantum computers. Finally, Muqeet \textit{et al.} \cite{MuqeetSSBSE2024} proposed an approach that employs genetic programming to develop noise models for quantum computers; such models can be used for testing quantum programs in simulators.

Though all the above work is contributing to good results in software testing, we can identify open challenges and opportunities that will need to be addressed in the next years, some of which are given below:

\begin{tcolorbox}[colback=LightGray, title=Challenges in Quantum Software Testing, colframe=DarkGray]
\begin{challenges}
    \item Efficient test oracles.
    \item Test scalability.
    \item From simulators to real quantum computers.
    \item (Quantum) Artificial Intelligence (AI) and (Quantum) Software Testing.
\end{challenges}
\end{tcolorbox}
\label{ch:st}

\textbf{Efficient test oracles.} Though some test oracles exist to assess whether test cases pass or fail, most of them are expensive to compute \cite{ali2021assessing,wang2021quito,StatisticalAssertions, ProjectAssertions,long2024equivalence}. In addition, sometimes the test oracle must be generated for each given input-output, which increases the execution cost of the testing process \cite{GarciaDeLaBarrera2022}. Given the scarce and expensive quantum computing resources, such test oracles do not scale. Thus, more efficient test oracles are needed. Theoretically, one could use state vectors instead of multiple executions, but one cannot easily obtain such state vectors from real quantum computers. Quantum state tomography and partial tomography can give us an approximation of the state vectors, but they are computationally expensive \cite{smolin2012efficient,bonet2020nearly}. In addition, there is also no efficient support (adding dynamic assertions is cumbersome) for checking quantum program states at various points during execution. Such checks, like in classical computing, are important, also for debugging and repair. Projection-based oracle \cite{ProjectAssertions} is one approach, but it still requires reading quantum programs, albeit it aims at reducing the times reading is required. Near future research directions include conducting empirical studies to devise guidelines for conducting experiments and selecting appropriate statistical techniques to keep the balance of required runs and the reliability of results. Another research direction aligns with classical test optimization: prioritizing testing on critical test cases. In this sense, the use of metrics and the development of quality models may be of great use in improving various aspects of the quantum software testing process \cite{Lemus2021,zhao2021some,Diaz-Munoz2024}. Additionally, exploring the conversion of testing tasks into common relations between quantum states or programs offers a valuable avenue. For example, many testing tasks can be transformed into checking the equivalence of two quantum programs or verifying whether the target quantum program is identity or unitarity \cite{long2024equivalence}. In this regard, discovering more topical relations between quantum states and programs would be valuable.

\textbf{Test scalability.} Most existing test data generation approaches are based on initial classical states of quantum programs (e.g., \cite{wang2021quito,qusbtICSE22}). QuraTest \cite{ye2023quratest}, and the work presented in \cite{long2024testing} goes beyond classical test input generation. However, the current works do not scale to complex quantum software and have challenges in efficiently covering the high-dimensional test input space determined by the number of qubits, entangled and superposition states, etc. Thus, we foresee advanced test data generation methods that lead to the efficient detection of faults in non-trivial quantum programs (e.g., faulty superpositioned states, wrong gates applied), considering practical constraints such as limited quantum computing resources and the diversity and representatives of the test data as we typically do in classical software testing. 

\textbf{From simulators to real quantum computers.} Currently, quantum computers are noisy; hence, test results are unreliable. Most current works test on ideal simulators (e.g., \cite{wang2021generating,qusbtICSE22,QuCat,Rui_metamorphic,wang2018quanfuzz,honarvar2020property,long2024equivalence,long2024testing,ye2023quratest}). Thus, we foresee the need for noise reduction techniques to be integrated as part of quantum software testing to increase the reliability of test results. Moreover, due to limited access to real quantum computers, existing testing solutions mainly rely on simulators. However, in the future, when real quantum computers become more accessible, it is crucial to test quantum programs on real quantum computers. Thus, to be cost-effective, test optimization would be needed \cite{Diaz2024}. Depending on the scale of testing, either classical test optimization or quantum test optimization approaches could be developed to minimize or prioritize tests. To this end, several preliminary works apply quantum annealing and quantum approximate optimization algorithms to optimize test cases for classical software \cite{wang2023test,testMinqaoaTSE2024}, which, we believe, is equally applicable to test optimization of quantum software.

    %\textbf{Further development of testing.} %Concerning mutation testing, different problems remain open. First, it is necessary to understand if artificial faults represent real faults. Empirical studies are also needed to discover existing subsumption relations among mutation operators due to the large number of mutants that can be produced. Further research is also needed to identify similarities and differences between classical and quantum software artifacts, guide the development of new tools or reuse existing ones, identify common pain points, and devise best practices through mining software repositories \cite{de2022software}, etc. Moreover, for quantum software, mapping testing techniques and activities to specific test phases is challenging \cite{miranskyy2021testing}. Formalizing this mapping for unit testing and integration testing has been performed \cite{long2024testing}, but other phases require more work.

\textbf{(Quantum) Artificial Intelligence (AI) and (Quantum) Software Testing.} There is a potential to use classical AI for quantum software testing and apply quantum AI to support classical software testing (for more details on the intersection between AI and quantum software engineering, also see the \hyperref[ch:ai]{\textit{Ch-AI}} challenges in Section \ref{sec:ai}). Classical AI techniques can support various quantum software testing and debugging. For example, there is potential to explore whether Large-Language Models (LLMs) can be used to assist in repairing quantum programs by providing patch recommendations \cite{guo2024repairing}. Moreover, LLMs can also be used during mutation analysis to obtain suggestions for generating realistic mutants for a given quantum program. Furthermore, classical AI techniques such as LLMs can be explored to help in determining appropriate test oracles and generating test cases by leveraging their ability to comprehend the semantics of quantum programs. For further discussion, see Section \ref{sec:ai}. \\ Quantum AI, on the other hand, has the potential to be used to empower classical software testing. For instance, quantum extreme learning machines have been explored to improve the efficiency of regression testing of industrial elevators compared to classical machine learning \cite{QUELL}. Quantum Approximate Optimization Algorithm (QAOA) has also recently been applied to solve test case optimization problems \cite{testMinqaoaTSE2024}, in addition to quantum annealing for classical test case minimization \cite{wang2023test}.  \\ These applications highlight the potential of applying quantum algorithms, in general, to improve various classical software engineering tasks (e.g., Grover's algorithm \cite{Grover_1998} has been utilized to speed up dynamic software testing for classical computers \cite{miranskyy2022using}). The key of such applications is to ensure that a practical software engineering task can be formulated into a mathematical problem that can be efficiently solved on a quantum computer \cite{miranskyy2022quantum}.  Numerous use cases span the entire software development lifecycle, from planning and requirements engineering to testing and maintenance \cite{miranskyy2022quantum}. 

\textbf{Other aspects of testing.} Additionally, other aspects of quantum software testing are worth discussing. Concerning mutation testing, different problems remain open. First, it is necessary to understand whether artificial faults really represent real faults. Empirical studies are also needed to discover existing subsumption relations among mutation operators due to the large number of mutants that can be produced. Further research is also needed to identify similarities and differences between classical and quantum software artifacts, guide the development of new tools or reuse existing ones, identify common pain points, and devise best practices through mining software repositories \cite{de2022software}, etc. Refactoring has been proposed to improve code maintainability in quantum software. Recent work introduced methods for applying refactorings specific to quantum programs, focusing on managing quantum-specific dependencies and optimizing circuit modularity \cite{zhao2023refactoring}. Moreover, for quantum software, mapping testing techniques and activities to specific test phases is challenging \cite{miranskyy2021testing}. Formalizing this mapping for unit and integration testing has been performed \cite{long2024testing}, but other phases require more work. To guide testing practices, seven testing principles have been proposed to guide quantum software testing \cite{long2024testing}. These principles emphasize representative and verifiable quantum input states, systematic partitioning of input variables by state types, and criterion-based input selection for comprehensive testing coverage. Additionally, ensuring clear methods for input preparation and output verification, maintaining endian mode consistency, and providing tools with key functionalities such as subroutine composition and result analysis are critical for effective testing. In addition, it is necessary to apply techniques that assist in the replicability and verifiability of quantum software experiments. To this end, there are already proposals for reporting guidelines and delineating a laboratory package structure adapted to quantum computing experiments \cite{Moguel2024Experiments}.

%%%%%%%%%%%%%%%%%%%%%%%%%%%%%%%%%%%%%%%%%%%%%%%%%%
\subsection{Service-Oriented Computing}
\label{sec:services}

% \note{REVIEWER COMMENT: it is not explicitly clear how traditional approaches to Service-Oriented Computing fail to apply to quantum software. Use of a concrete example application may help illustrate why this challenge is important.}

In classical software engineering, Service-Oriented Computing (SOC) is defined as a paradigm to support the development of interoperable, distributed applications where services are regarded as autonomous entities \cite{papazoglou2007service}. In the last few years, research in this field has covered a range of topics, including the provisioning and aggregation of basic services \cite{Valencia2021}, the composition of complex systems through the coordination of services \cite{AlvaradoValiente2024Balancing}, and the management and monitoring of services deployed in the cloud or on-premise \cite{RomeroAlvarez2023}. 

Quantum hardware and software offered in the cloud allow users to access quantum computing resources over the internet without the need to own quantum computing hardware. Several companies and institutions provide cloud-based quantum computing services like IBM already mentioned above, or others like Amazon Braket or Microsoft through Azure Quantum. This makes it easier for a broader audience to explore and develop quantum software. These platforms offer access to a variety of quantum processors and simulators, along with tools and environments to create, run, and test quantum algorithms.

%In this scenario, it might appear that using classical SOC principles is easy. 
Employing classical SOC principles in this scenario may look easy. Nevertheless, given the incipient state of quantum hardware technology, classical SOC approaches are not directly transferable to quantum software development \cite{RomeroAlvarezBook2024}. 
The distinct and complex interaction model required for quantum computing challenges the straightforward application of conventional SOC frameworks, so it is necessary to re-evaluate how these principles can be adapted or redefined for the field of quantum computing \cite{Moguel2022}. To address these difficulties, several research groups have recently focused on Quantum Service-Oriented Computing, intending to bring its benefits to Quantum Software Engineering \cite{Moguel2020}. 

% \note{REVIEWE COMMENT: Use of citations as nouns, e.g., “in [131]”, is not grammatically correct. It is better to have something like “Author presented one of the first approaches [131]”}

Wild \textit{et al.} \cite{wild2020tosca4qc} presented one of the first approaches in this regard. In this work, the authors presented TOSCA4QC, introducing two deployment modeling styles based on the Topology and Orchestration Specification for Cloud Applications (TOSCA) standard for automating the deployment and orchestration of quantum applications. From there, this research group has provided several contributions in the Quantum SOC domain. Also, the work done by Weder \textit{et al.} \cite{weder2021hybrid} focuses on the need for additional mechanisms for orchestration in hybrid quantum applications over classical applications. On the other hand, Beisel \textit{et al.} \cite{beisel2022quokka} provide a service ecosystem for the execution, based on workflows, of Variational Quantum Algorithms (VQA). VQA is a hybrid quantum-classical approach that uses parameterized quantum circuits and classical optimization techniques to solve complex optimization and machine learning problems on near-term quantum devices \cite{cerezo2021variational}. 

Also, Weder \textit{et al.} \cite{weder2023provenance} introduce a method to detect parts of a hybrid workflow that can be optimized at runtime. These works, as a whole, provide a formalized and complete toolkit to bring the benefits of working with workflows for the deployment and execution of hybrid quantum applications.

From a different perspective, the work done by Rojo \textit{et al.} \cite{rojo2021trials} analyzes the problems and difficulties of developing hybrid quantum applications using a service-oriented approach. From there, this research group has also provided several contributions in this domain. Also, Garcia-Alonso \textit{et al.} \cite{garcia2021quantum,AlvaradoValiente2024} proposed adapting the API Gateway pattern to be combined with quantum software. And Romero \textit{et al.} \cite{romero2022using} extend the Open API standard to develop quantum services. Another work by Romero \textit{et al.} \cite{romero2023enabling} focuses on using DevOps techniques to enable the continuous deployment of quantum software. Finally, Bisicchia \textit{et al.} \cite{bisicchia2023distributing} propose a technique for distributing quantum computations between different computers by splitting the shots required for a given quantum task among different QPUs. All these works provide an approach that brings the development of hybrid quantum applications closer to the tools and techniques usually employed in SOC.

These examples show the potential benefits of applying the principles of Service-Oriented Computing to Quantum Software Engineering. However, to exploit the full potential of service orientation, several challenges must be addressed first. More specifically: 

\renewcommand{\arealabel}{SoC}
\begin{tcolorbox}[colback=LightGray, title=Challenges in Quantum Service-Oriented Computing, colframe=DarkGray]
\begin{challenges}
    \item Interoperability.
    \item Platform independence.
    \item Demand and Capacity Management.
    \item Workforce training.
\end{challenges}
\end{tcolorbox}
\label{ch:soc}

% \textbf{Interoperability.} As different technologies and platforms for Quantum Computing keep emerging, the need for industry standards or proposals of such made by researchers becomes clear. Some examples exist, like OpenQASM standardized by several companies, that showcase the advantages of such standardization. But there are still various assembly-level languages, thus efforts for creating some kind of intermediate format (like QIR) are needed in order to progress with this challenge. Similarly, the creation of standard APIs to interact with QPUs will significantly improve interoperability and pave the way for researchers to create a new generation of quantum service-oriented computing tools and solutions.

\textbf{Interoperability.} Facing this challenge is crucial as the quantum computing landscape continues to expand, with various technologies and platforms emerging across the industry. The lack of standardized frameworks hinders seamless interaction between different QPUs, software stacks, and development environments. The need for universally accepted standards, either through industry initiatives or research-driven proposals, is becoming increasingly evident. 
    
One notable example is OpenQASM (Quantum Assembly Language), standardized by several companies, highlighting the benefits of unifying quantum programming languages. OpenQASM allows developers to write instructions compatible with multiple quantum computers, demonstrating how standardization can simplify the development process and reduce fragmentation. However, despite this advancement, there are still numerous assembly-level languages in use, each tied to specific hardware or platforms, creating a barrier to widespread quantum software development. 
    
To address this challenge, ongoing efforts are focusing on creating intermediate representations, such as the Quantum Intermediate Representation (QIR), which can act as a bridge between high-level programming languages and low-level hardware-specific instructions. QIR allows quantum programs to be hardware-agnostic, enabling code to be executed across different quantum platforms without the need for reimplementation. This not only facilitates portability but also enhances scalability as quantum hardware evolves. 
    
In addition to programming languages, there is a pressing need to create standardized APIs to interface with QPUs. A consistent set of APIs would enable developers to build quantum applications without needing to understand the specifics of each quantum device deeply. This would significantly improve the interoperability of quantum systems and enable the development of tools and solutions aligned with quantum service-oriented computing. Standard APIs could allow services to interact seamlessly across different QPU architectures, fostering collaboration and allowing for more sophisticated quantum cloud services, cross-platform quantum software development, and distributed quantum computing. By establishing these standards, the quantum computing community can ensure that future quantum technologies are accessible, interoperable, and scalable across diverse platforms.

\textbf{Platform independence.} This challenge is a key issue in the rapidly evolving quantum computing ecosystem, where the capabilities of QPUs vary widely across different vendors. The environments through which these QPUs are accessed (often via cloud-based platforms) introduce further variability. Some platforms offer hybrid runtimes that integrate classical and quantum computing, while others provide software optimizations and performance enhancements that are specific to their own architecture. This diversity in features, while beneficial in some ways, creates challenges for quantum software developers, who must often tailor their applications to specific platforms. As a result, there is a significant dependency on the platform for which the software is designed, increasing the risk of vendor lock-in where developers are constrained to a particular hardware provider, limiting flexibility and innovation.
    
In classical service-oriented computing, platform independence has been a driving force for the scalability and success of cloud-based services. By abstracting the underlying hardware, developers can deploy applications on a variety of systems without having to modify their code for each platform. Achieving a similar level of platform independence in quantum computing is crucial to fostering a more open and competitive ecosystem where developers can choose the best tools and services without being tied to a specific vendor’s hardware or cloud environment.
    
To realize platform independence in quantum computing, intermediate layers that abstract away the details of specific QPUs are needed. This could involve the development of universal quantum compilers that translate high-level quantum programs into intermediate representations compatible with multiple hardware platforms. Standards such as QIR (already mentioned above) as well as efforts to create vendor-neutral quantum software development kits (SDKs) could serve as key building blocks for this. These SDKs would enable developers to write quantum applications that run across various quantum hardware without the need for significant reconfiguration or optimization.
    
    % Furthermore, quantum cloud services should evolve to support platform-independent development environments, offering APIs and tools that work seamlessly across different quantum architectures. This would not only lower the barrier to entry for developers by reducing the need for specialized knowledge of each platform but would also encourage the creation of quantum service-oriented architectures. In such architectures, developers could build applications that dynamically leverage the strengths of various quantum and classical resources, making use of hybrid systems while maintaining flexibility across platforms.
    
    % Achieving true platform independence in quantum computing will require significant collaboration between hardware vendors, software developers, and standardization bodies. However, the payoff would be substantial—allowing quantum software to reach its full potential by enabling interoperable, scalable, and flexible quantum solutions that can be deployed across a variety of quantum hardware environments. This would reduce the risk of vendor lock-in, foster innovation, and allow the quantum ecosystem to grow in a more open and accessible way.

\textbf{Demand and Capacity Management.} This challenge in quantum computing is complex, particularly when supporting hybrid workflows that integrate both classical and quantum resources. Effective coordination between these systems is essential, as quantum computations often rely on classical pre-processing and post-processing steps. Additionally, optimizing data transfer and communication between classical and quantum systems is crucial for maintaining efficiency, as quantum computers may be physically remote and subject to network latency or bandwidth constraints. These factors make it imperative to design infrastructure that seamlessly manages the interaction between classical and quantum resources to avoid bottlenecks. 
    
Quantum hardware introduces further complexity into capacity management due to its inherent limitations, such as coherence times (the duration for which a qubit remains in a superposition), gate fidelities (the accuracy of quantum operations), and qubit connectivity (the ability of qubits to interact with one another within a quantum processor). Capacity planning for quantum services must consider these constraints, which directly impact the performance and scalability of quantum computations. The number of available qubits, their quality, and the complexity of quantum circuits that can be executed in a single run are key factors that must be continuously monitored and optimized. 
    
A significant challenge lies in adapting existing demand and capacity management strategies, particularly from microservice architectures, to quantum computing. In traditional microservice-based systems, capacity is often analyzed and managed with third-party services in mind \cite{fresno2022}. In scenarios where these services are quantum, however, new considerations emerge—such as the limited availability of quantum processing time, the need for queuing systems to manage access to quantum hardware, and the potential cost implications of quantum computing resources. For example, the latency and reliability of third-party quantum services will differ drastically from classical services, necessitating new models for assessing service-level agreements (SLAs) and ensuring that capacity is dynamically allocated to meet fluctuating demand. 
    
Moreover, the role of pricing plans \cite{FresnoAranda2025} and capacity limitations in the API industry \cite{gamez2019ESEC} will take on new significance as quantum services become commercially available. Pricing models for quantum computing are typically based on the amount of time spent on a quantum processor, the number of qubits used, and the complexity of the quantum circuits. These factors introduce new challenges for capacity management, as developers must balance computational requirements with cost efficiency. As quantum APIs become more integrated into broader service architectures, organizations will need to develop sophisticated pricing and capacity strategies that account for the unique characteristics of quantum computing. This will likely include tiered pricing models based on quantum hardware capabilities, the development of more granular usage metrics, and dynamic capacity allocation based on real-time demand. 
    
    % In summary, managing demand and capacity in the quantum computing domain requires an approach that considers both the traditional aspects of hybrid workflows and the unique challenges posed by quantum hardware limitations. Existing capacity management frameworks, such as those used in microservices, must be adapted and expanded to accommodate the specific needs of quantum services. Additionally, as quantum computing becomes more integrated into service architectures, pricing models and capacity management strategies will need to evolve, ensuring that quantum resources are efficiently utilized while remaining cost-effective.

\textbf{Workforce training.} This challenge is critical as the industry transitions towards quantum and hybrid software development. As such, this challenge could be included in many other research areas discussed in this section. It has been included here due to the pervasive presence of service-oriented computing in industry projects. While the principles of service-oriented computing are well-understood and widely practiced by a large community of developers, introducing quantum computing into this paradigm requires a substantial shift in both mindset and skills. Developers accustomed to classical Service-Oriented Architecture (SOA) must be retrained to navigate the complexities of Quantum Service-Oriented Computing (QSOC). The challenge here is multifaceted, requiring not only new technical knowledge but also the ability to integrate quantum principles with existing classical architectures. 
    
Addressing this challenge involves a two-pronged approach. The first step is to define comprehensive training strategies that facilitate the smooth transition from classical to quantum development. These strategies should include specialized educational programs, certifications, and hands-on experience with quantum technologies. Training initiatives must be designed to bridge the gap between classical and quantum computing by focusing on core quantum concepts such as qubits, superposition, entanglement, and quantum algorithms while also teaching developers how these concepts integrate with classical computing frameworks. Moreover, training should be structured progressively, beginning with foundational quantum mechanics and programming languages like Qiskit or Cirq and moving towards more advanced topics like hybrid quantum-classical workflows and quantum cloud services. The end goal is to make quantum technologies accessible to traditional developers, easing the learning curve and fostering broader adoption across industries. 
    
The second aspect of addressing workforce training is the development of Quantum Software Engineering methodologies and tools that streamline the transition from classical Service-Oriented Computing to its quantum counterpart. These methodologies should provide a systematic framework for building, testing, and deploying quantum services within hybrid systems. For instance, tools that abstract the complexities of quantum hardware and offer familiar service-oriented interfaces will be crucial in helping developers adopt quantum technologies without needing deep expertise in quantum mechanics. By focusing on interoperability between classical and quantum components, these tools can reduce the cognitive load on developers, allowing them to build quantum-enhanced applications with minimal disruption to their existing workflows. 
    
Additionally, collaborative environments and quantum development platforms should be created to foster hands-on learning. Quantum simulators, cloud-based quantum platforms, and integrated development environments (IDEs) that support hybrid systems will be key to giving developers practical experience with quantum technologies. These platforms can provide sandbox environments where developers can experiment with quantum services, test quantum circuits, and integrate quantum capabilities into classical service-oriented applications. 
    
Beyond technical training, there must also be an emphasis on reshaping the development culture to accommodate quantum thinking. In classical service-oriented computing, concepts like modularity, scalability, and interoperability are well established, but in quantum computing, additional considerations such as quantum error correction, decoherence, and probabilistic outcomes are crucial. Developers need to be educated not only in the technical aspects of quantum computing but also in the unique challenges and opportunities it presents. Workshops, conferences, and community-driven learning initiatives can be valuable in fostering a collaborative culture where knowledge sharing and continuous learning are prioritized.
    
    % Ultimately, workforce training in the quantum domain is not a one-time initiative but an ongoing process. As quantum hardware and software technologies evolve, so too will the requirements for developers working with these systems. Therefore, continuous education and lifelong learning programs will be necessary to ensure that developers remain up-to-date with the latest advancements in QSE and quantum technologies. By investing in both education and tool development, the industry can prepare the current and next generation of developers to effectively harness the power of quantum computing within the service-oriented architecture paradigm.

%%%%%%%%%%%%%%%%%%%%%%%%%%%%%%%%%%%%%%%%%%%%%%%%%%
\subsection{Model-Driven Engineering}
\label{sec:MDE}

Model-Driven Engineering (MDE) \cite{schmidt2006model} has been historically significant in the development of classical software by providing methodologies and tools to manage complexity through high-level abstractions and automation. MDE involves creating abstract models that represent the system's structure, behavior, and functionality, allowing for a clear and concise specification of requirements and design. These models are often defined using Domain-Specific Modelling Languages (DSMLs) tailored to particular domains, enhancing expressiveness and reducing ambiguity. MDE facilitates model validation and verification through simulation and formal methods, ensuring that the software adheres to the specified requirements before implementation. One of MDE's key strengths is automatic code generation, where high-level models are transformed into executable code, reducing manual coding effort and minimising errors. This automation extends to various phases of software development, including analysis, design, implementation, and testing. By promoting a model-centric approach, MDE improves the consistency, traceability, and maintainability of software systems, enabling better alignment with business goals and more efficient handling of complex large-scale projects.

As can be seen, MDE is transversal to the other sections and topics addressed in this article, as it can contribute significantly to all areas. However, this section focuses more specifically on abstract modelling and automatic transformation of quantum and hybrid software models.

One of the most relevant challenges of current QSE is that existing languages and methodologies in quantum software development operate at a notably lower level of abstraction compared to the typical scope addressed by MDE techniques for classical software. The fundamental differences between classical and quantum software complicate the development and reuse of MDE techniques for this kind of application (e.g., boolean algebra vs. quantum mechanics principles, well-known architectures and patterns vs. algorithm complexity, deterministic vs. stochastic, low dependency and high compatibility with hardware vs. the opposite, high scalability vs. NISQ limitations). Furthermore, the design of hybrid software is rarely modelled at a higher abstraction level, ignoring irrelevant low-level technical details and focusing on architectural details (see Section \ref{sec:architecture}). To address these challenges, several research groups have started working on Model-Driven Quantum Software Engineering and have already produced relevant contributions in the area.

Ali and Yue \cite{AliTao-MDE} offer a preliminary exploration of how MDE can be used to support quantum code generation or quantum verification and validation. Similarly, Gemeinhardt \textit{et al.} \cite{gemeinhardt2021towards} propose an initial roadmap of research questions that should be addressed to bring the benefits of MDE to quantum software. From this starting point, the authors have developed several additional contributions. Gemeinhardt \textit{et al.}. \cite{gemeinhardt2023model}, in a different work, analyze how model-driven optimization techniques can be applied in the context of quantum software. Furthermore, the same authors propose a modelling language and a design framework for quantum circuits that support the definition of composite operators \cite{wille2024model}. This allows developers to raise the abstraction level of quantum algorithm design alongside the provided code generator. Similarly, Ammermann \textit{et al.} \cite{ammermann2024mde} propose a view-based quantum development approach based on a Single Underlying Model (SUM). That proposal is supported by a quantum Integrated Development Environment (IDE) to model quantum software at a higher abstraction level by considering different specific views for various stakeholders.

With a different goal but using similar model-driven principles, software modernization efforts have been tailored to tackle the issues associated with migrations of these hybrid software systems \cite{perez-castillo-2021}. The software modernization process, which integrates traditional reengineering with MDE principles, has been progressively developed by Pérez-Castillo \textit{et al.} over the last few years. Thus, reverse engineering techniques to abstract different quantum programming languages have been proposed \cite{perez2022qrev}; the restructuring and transformation of different high-level representations have been addressed \cite{jimenez2021kdm}; or preliminary code generation techniques from high-level designs of hybrid software have been proposed \cite{perez2023-egl}.

One of the most recurrent challenges covered is how to model quantum/hybrid software in an abstract way. In this sense, Pérez-Delgado and Perez-Gonzalez \cite{PerezDelgadoUML}, outlined certain principles for designing modelling languages for quantum software. Similarly, Pérez-Castillo \textit{et al.} \cite{UMLProfile}, propose a UML profile that covers the analysis and design of hybrid software. In addition, Ali and Yue \cite{AliTao-MDE} discuss some ideas for obtaining new metamodels for modelling quantum programs as extensions of UML. Apart from UML, other authors have focused on other existing standards. Weder \textit{et al.} \cite{weder2020integrating} introduce a BPMN-based modelling approach to facilitate the integration of quantum computations with classical applications and quantum circuits, aiming to simplify orchestration tasks and ensure portability. Zhao \cite{zhao2024towards} proposes a foundation for developing architecture description languages (ADLs) specifically designed for hybrid quantum-classical software systems. The research aims to establish a formal framework for describing the architecture of such systems by capturing both quantum and classical components and their interactions at the architectural level. In addition, some of the works mentioned \cite{perez2022qrev, jimenez2021kdm} use the extension of the Knowledge Discovery Metamodel (KDM) to support the maintainability of quantum software.

In contrast to extensions for existing modelling standards, some authors have explored using DSMLs, as discussed by Gemeinhardt \textit{et al.} \cite{gemeinhardt2021towards}. DSMLs, such as SimuQ \cite{Peng2024} and Quingo \cite{Fu2021}, cater to specific needs within quantum computing. Polat \textit{et al.} \cite{polat2024mde} provide MDE4QP, a framework built upon the existing MDE tools in classical platforms with which to define optimisation problems in a Platform-Independent Model (PIM), and then provides automatic transformations for different Platform-Specific Models (PSM) such as gate-based or annealing programming models.

Another relevant DSML is Quantum Intermediate Representation (QIR), which has been developed by Microsoft. \cite{geller2020introducing}, QIR serves as a DSML built on the LLVM intermediate language, to provide a unified interface between quantum programming languages and platforms. There exist other similar libraries and frameworks that make the code hardware-independent. For example, Xanadu Pennylane software and, to some extent, AWS Braket software can convert their code to multiple hardware vendors. Finally, other examples of DSMLs have been proposed for Quantum Machine Learning \cite{Moin2021} and for modelling quantum circuits derived from satisfiability problems \cite{Alonso2022}.

Those are only some examples of the work in the intersection between MDE and Quantum Software Engineering. However, additional research efforts are still needed to address the remaining challenges in this domain. Specifically:

\renewcommand{\arealabel}{MDE}
\begin{tcolorbox}[colback=LightGray, title=Challenges in Quantum Model-Driven Engineering, colframe=DarkGray]
\begin{challenges}
    \item Modelling quantum-specific constructs.
    \item Development of high-level design methodologies.
    \item Scalable quantum software maintenance and evolution.
    \item Intelligent code generation and orchestration.
\end{challenges}
\end{tcolorbox}
\label{ch:mde}

\textbf{Modelling quantum-specific constructs}. An open challenge for the future in the application of MDE to quantum software lies in developing quantum-specific modelling constructs. Unlike classical software, quantum software requires precise representations of quantum gates, circuits, and states, which necessitate specialised MDE languages, techniques, and tools capable of accurately modelling these components. Additionally, the inherently probabilistic nature of quantum computations, including aspects such as measurement probabilities, must be effectively incorporated into MDE models. In this sense, there already exists some approximations for modelling uncertainty \cite{troyaUncertainty}, which should be further explored for quantum software. Addressing these challenges is crucial for advancing the reliability and accuracy of quantum software development using MDE methodologies.

\textbf{Developing high-level design methodologies for hybrid software systems} is crucial for bridging the gap between classical and quantum computing paradigms. This involves creating abstract modelling frameworks that encapsulate the complexity of hybrid quantum-classical interactions, providing a unified view that enhances comprehensibility and facilitates design decisions \cite{zhao2024towards}. Future research could focus on developing DSMLs that offer intuitive abstractions for quantum-classical integration, enabling software engineers to design hybrid applications without delving into the low-level technical aspects of quantum computing.
    
\textbf{Scalable quantum software maintenance and evolution.} As quantum software becomes more complex and widespread, maintaining and evolving these systems will pose significant challenges. Future research could explore MDE approaches to predict the impact of changes in quantum software components also involving sophisticated quantum software metric models, ensuring compatibility and optimising performance across versions. Techniques such as model-based regression testing and automated refactoring tools tailored to quantum software could be developed to support scalable maintenance processes.
    
\textbf{Intelligent code generation and orchestration} are crucial to improving the productivity and efficiency of QSE. By automating the generation of quantum code from high-level models, developers can focus more on problem-solving rather than the intricacies of quantum programming languages. Future work in this domain could look at developing sophisticated code generation engines, e.g., based on model-driven optimisation, that translates models into executable quantum code and optimises this code for specific quantum hardware, considering factors such as qubit connectivity and gate fidelity. Orchestration, on the other hand, involves managing the execution of quantum and classical components in hybrid systems, ensuring they operate seamlessly together to achieve desired outcomes. Furthermore, the research could aim to create intelligent orchestration tools that dynamically manage the execution of hybrid applications, optimising resource allocation and execution in order to improve performance and reliability.

%%%%%%%%%%%%%%%%%%%%%%%%%%%%%%%%%%%%%%%%%%%%%%%%%%
\subsection{Programming Paradigms}
\label{sec:paradigms}

Programming consists of instructing computers with algorithms (strategies) to achieve an objective (produce a result) using programming languages \footnote{By programming language here we mean any special or dedicated language (including graphical or natural languages) with enough precision to express algorithms that finally can instruct computers.}. The typical approach to formulating strategies involves breaking them down into increasingly simpler steps until the level of instruction provided by programming languages is reached. The process of breaking down occurs at different levels of abstraction. Thus, the way strategies are approached is conditioned by the final set of operations available for the computation model, be it classical or quantum.

Since 1996 to date, when the metalanguage lambda-q calculus was proposed \cite{maymin1997extending}, many quantum programming languages have been created. Most of them are languages that follow the imperative programming paradigm. Some examples of them are QASM \cite{qasm} based on Assembly, Ket \cite{ket} based on Python or Q\# \cite{Qsharp} based on C\#. A few of them are functional, like QML \cite{QML} and Quipper \cite{green2013quipper} based on Haskell, or even declarative languages, like Forest \cite{forest} based on Python.

All of the above languages are designed to compose quantum circuits that will compute on a QPU. Programming circuit-based quantum computers presents a significant challenge for classical programmers approaching it \cite{QOPNeed}. While many factors contribute to it, we focus here on two of them. 

On the one hand, the conception of strategies for classical and quantum programs differs significantly. A classical program encodes a strategy for composing a result\footnote{Strategies to produce results are not only for imperative languages. Logic or declarative languages also express strategies to produce results with instructions conceived for their different paradigms}. Each step in a classical program transforms the machine's state. Eventually, the result is produced, and the program is deemed correct because it successfully constructs the result. On the contrary, a quantum program encodes a strategy for discovering the result. 
%In a quantum program, computations typically begin by generating a superposition state in which the solution(s) already exist in one or several configurations of the state. The strategy consists of a sequence of steps to amplify the amplitude of those configurations.
In a quantum program, calculations usually begin with initialisation operations on a quantum register, thus generating a quantum state in which different values coexist in superposition. Each of these values has an associated probability amplitude collected in the corresponding state vector associated with the state. The program ends by collapsing the quantum state. The resulting value after collapse is one of the values of the quantum state in superposition. Therefore, this result already exists in the quantum state. The program strategy is then determined by all manipulations of the quantum state after its initialization and until its collapse. Such manipulations are aimed at increasing the probability amplitude of the value(s) that constitute a solution to the problem solved by the program.

On the other hand, quantum programming languages provide a low level of abstraction. Although there are many languages for quantum programming, such as QASM, Quil, qibo, and Qiskit, all of them constrain the abstraction level to that provided by the primitives of the language (quantum gates). Those primitives operate directly over the phase and amplitude of qubits representing the quantum state. So, quantum program strategies must be considered in terms of phase and amplitude manipulation. Compared to classical programming, it is like composing strategies in terms of rotation, shifting, addition, or carrying operations over binary registers. This stage has long been superseded in classical software engineering, and history has shown its benefits.

In recent years, some initiatives have been developed that seek to make quantum programming more affordable by raising the level of abstraction and providing abstractions that facilitate the design of discovery strategies. One of the significant advances in achieving comprehensibility, reliability, and simplicity of classical software was the introduction of basic data types like Integer, Float, or Character, along with simple operations on them. Now, inspired by that \cite{10189119}, \cite{less2} and \cite{10.1007/978-3-031-45728-9_7} propose a kind of Oracle Based Quantum Programming. The idea is to treat quantum registers as data-type encodings. Such types are complemented with oracles  \cite{typesOracles2019}, \cite{Leymann-QuantumAlgorithms} that implement basic operations on them. The feasibility of this approach has been explored considering quantum registers encoding integers. Then, a set of reusable and composable oracles implementing simple operations \cite{10.1007/978-3-031-45728-9_7} have been developed. 

To provide quantum programmers with a higher abstraction level, other authors also exploit the idea of providing Quantum Types \cite{varga2024quantum}. Programmers can define their own types and thus create new abstractions on which to build their algorithms. Some other works are also focused on providing different means to encode specific types and operations in quantum states \cite{Wiebe_2013, cuccaro2004new, häner2018quantum, seidel2021efficient}

The concern of raising the level of abstraction to make the programming of quantum systems more accessible is not unique to circuit-based systems. Also, researchers in the field of annealing quantum computing aim to achieve this objective. Programming a QUBO is not an easy task. Thus, some researchers \cite{Codognet} try to approach that problem by enabling programmers to formulate their algorithms in the scope of Constraint Programming \cite{mayoh2013constraint} and then translating constraints to a QUBO. 

The above work reveals some of the challenges that will have to be addressed over the next years:

\renewcommand{\arealabel}{PP}
\begin{tcolorbox}[colback=LightGray, title=Challenges in Quantum Programming Paradigms, colframe=DarkGray]
\begin{challenges}
    \item Complexity of circuits.
    \item Composable and reusable quantum software.
    \item Abstractions for quantum software.
\end{challenges}
\end{tcolorbox}
\label{ch:pp}

\textbf{Complexity of circuits.} This is one of the core challenges in quantum computing, particularly when it comes to implementing algorithms that require sophisticated operations. Oracles, which serve as black-box subroutines, have shown their potential as a pattern for encapsulating complex operations. These oracles are essential in many quantum algorithms, such as Grover's search algorithm, where the amplitude amplification technique is encoded as a subroutine that can be reused in larger computations \cite{Grover_1998}. The ability to encapsulate such operations is invaluable because it abstracts the complexity, allowing developers to focus on higher-level algorithmic design. However, as with any complex quantum operation, implementing these oracles can result in circuits that are both wide (involving many qubits) and deep (requiring numerous sequential gate operations), which are challenging for current quantum hardware.
    
The wide and deep nature of these circuits presents serious constraints, particularly with NISQ devices, which are limited by coherence times, gate fidelities, and qubit connectivity. A wide circuit implies a large number of qubits, which is difficult to achieve with current hardware, while deep circuits require a high number of gates, which increases the likelihood of errors due to noise and decoherence. Moreover, deep circuits can be inefficient for near-term quantum computers, as long-running computations exceed the coherence time of qubits, leading to a loss of quantum information. Consequently, optimizing quantum circuits to minimize both the number of qubits and the number of sequential gate operations is crucial for making complex algorithms feasible on today's hardware.
    
Researchers are actively exploring optimization techniques to address these challenges. One promising direction involves quantum gate synthesis, where gate sequences are optimized to reduce circuit depth \cite{Iranmanesh2024}. Similarly, techniques such as circuit compression aim to minimize the number of gates required for a given quantum operation, while qubit recycling strategies reuse qubits within a circuit to reduce the overall qubit count \cite{Sano2024}. Another approach is hardware-aware optimization, where the specific architecture of a quantum device, such as qubit connectivity or gate fidelity, is considered during the design of the circuit \cite{Niu2020}. This allows circuits to be mapped onto hardware in a way that reduces the need for costly operations like SWAP gates, which arise from poor qubit connectivity. These advancements in circuit optimization are critical for the practical implementation of oracles and other complex quantum subroutines on real-world devices.

\textbf{Composable and reusable quantum software.} This challenge is essential for reducing the complexity of designing quantum algorithms and circuits. Developing oracles or other complex quantum circuits often requires a deep understanding of quantum mechanics and quantum states, making the process daunting for many developers. To make quantum development more accessible and efficient, it's crucial that these circuits are designed to be as reusable as possible. By creating quantum circuits that can be reused across multiple algorithms and applications, the overall effort required to develop new quantum solutions is significantly reduced. Reusability in quantum software not only saves time and resources but also enables developers to build upon existing, well-optimized circuits rather than starting from scratch each time a new algorithm is needed.
    
For circuits to be reusable, they must also be composable, meaning they can be integrated with other circuits seamlessly. This is where the current landscape of quantum development faces a challenge. The tools and frameworks available to support circuit composition and reuse are still in their infancy, lacking the sophistication seen in classical software engineering. In classical computing, developers have access to mature libraries, modular components, and standardized interfaces that enable the efficient reuse and integration of code. In contrast, quantum computing is still developing such frameworks, and the process of reusing quantum circuits is far more complex due to the unique nature of quantum states, entanglement, and superposition. Furthermore, quantum circuits often need to maintain delicate quantum properties like coherence and phase relationships, which adds additional constraints when trying to compose them with other circuits.
    
Initial efforts to create tools and techniques for the documentation, reuse, and composition of quantum software have highlighted just how different the quantum domain is from classical software engineering \cite{reusing}. Unlike classical software, where code can be modularized and reused with relative ease, quantum circuits require more specialized handling to preserve their quantum characteristics when integrated with other circuits. This makes the task of developing reusable quantum components much more challenging. Over the next years, the development of techniques for better documenting quantum circuits, creating modular quantum software architectures, and enabling the flexible composition of quantum components will be crucial. Such advancements will help quantum developers reuse complex quantum logic efficiently, thereby accelerating the development of new quantum applications and making quantum software development more scalable and accessible for a broader range of developers.

\textbf{Abstractions for quantum software.} This challenge is essential for elevating the usability and flexibility of quantum programming languages. At present, the process of encoding data types in quantum states and designing corresponding operations on these states is one of the foundational strategies for increasing the abstraction level in quantum computing. This strategy opens up the potential for defining new sets of operations that enrich the primitive operations currently provided by quantum programming languages. These operations, in turn, can offer developers more versatile and powerful tools for designing quantum algorithms. However, the types and operations that have been proposed so far in the literature often mirror those used in classical computing, limiting the scope of what quantum computers can achieve. For instance, some researchers have attempted to define quantum states that encode prime numbers \cite{primes}, but these kinds of abstractions, while useful, might not fully exploit the unique capabilities of quantum systems.
    
Quantum computing's true potential lies in its ability to model and simulate complex quantum systems, such as those found in chemistry, physics, and other fields where quantum effects dominate. Therefore, a more promising direction for developing quantum abstractions would be to focus on encoding quantum-native data types that reflect the kinds of problems quantum computers are inherently suited to solve. For example, instead of abstracting classical concepts like numbers or strings, quantum states could be used to represent more complex entities like molecules, atomic structures, or even quantum fields. Operations on these quantum states could simulate interactions between molecules or particles, allowing quantum algorithms to directly engage with the types of calculations that classical computers struggle with, such as simulating chemical reactions or quantum systems with many-body interactions. This approach would not only enhance the level of abstraction in quantum programming but also align more closely with the strengths of quantum computing as envisioned by pioneers like Richard Feynman \cite{feynman2018simulating}, who famously proposed that quantum computers would excel in simulating physical processes.
    
    % To date, the quantum computing research community has made limited progress in developing such domain-specific abstractions for quantum software. While some early efforts have focused on quantum machine learning or quantum cryptography, which abstract certain tasks into quantum operations, more general abstractions for fields like chemistry or materials science have yet to be fully realized. For instance, a quantum state representing a molecule that can interact with other molecular states through quantum operations could revolutionize how we simulate and understand chemical reactions or even drug discovery processes. Despite the promising potential, no well-defined quantum abstractions for these kinds of data types have been widely adopted, leaving a significant gap in the development of quantum software. Moving forward, it is crucial for the research community to invest in creating these higher-level abstractions, which will not only simplify the design of quantum algorithms but also unlock the full computational power of quantum systems in solving real-world problems.

%%%%%%%%%%%%%%%%%%%%%%%%%%%%%%%%%%%%%%%%%%%%%%%%%%
\subsection{Software Architectures}
\label{sec:architecture}

% \note{REVIEWER COMMENT: in the Architecture section, examples of design patterns proposed for quantum/hybrid software would be very illuminating}

% \note{REVIEWER COMMENT: The Architecture section is very short and undetailed, especially in comparison to the other sections. I would like to see more expansion and discussion of the challenges identified in this area.}

Software architecture involves creating structures essential for understanding and creating a software system. These structures consist of software elements, their relationships, and their properties. We can use quantum computers to make our classical software tackle problems that were previously out of reach. Consequently, quantum systems should not operate independently but should co-exist and collaborate with classical systems \cite{SEI2021,perez-castillo-2021,weder2021hybrid,zhao2024towards}. 

Tools and methodologies are needed to integrate quantum layers and stakeholders with those in classical information systems. The motivation for such integration of different systems should not just be the fact that they reside in different computational paradigms. Instead, information systems should be able to be designed as a whole and integrated based on the functionalities they provide independently of the hardware architectures in which they reside. 

% \note{REVIEWER COMMENT: Use of citations as nouns, e.g., “in [131]”, is not grammatically correct. It is better to have something like “Author presented one of the first approaches [131]”}

Software architects play a crucial role in achieving seamless integration while designing systems that effectively meet businesses' requirements. Thus, the systematic review by Khan \textit{et al.} \cite{Khan23} investigates the software architecture for quantum computing systems. Further challenges and opportunities are discussed in detail by Yue \textit{et al.} \cite{Yue2023}. In a similar study, aspects involved in various software architectures are analyzed since ``\textit{the software architecture of quantum computing systems plays a pivotal role in determining their ultimate success and usability}'' \cite{Zhao24}. In the work ``\textit{Architecture Decisions in Quantum Software Systems}'' \cite{aktar2023architecture}, the authors conduct empirical research to examine and analyze architectural decisions while creating quantum software systems. A foundation for developing ADLs has also been discussed by Zhao \cite{zhao2024towards}.

Some particular design patterns have also been proposed by Frank Leymann \cite{Leymann-QuantumAlgorithms} and Buhler \cite{buhler2023patterns}, which mainly focused on the design of quantum circuits. In the study \cite{perez-castillo-24-patterns}, authors perform a preliminary exploration of the usage in the practice of some of those design patterns. Preliminary work in extending the definition of patterns for hybrid software systems is proposed in \cite{weigold2021patterns}. Guo \textit{et al.} \cite{guo2024quantum} specifically focused on quantum circuit ansatzes, a type of specialized design patterns commonly used for variational quantum algorithms (VQAs) \cite{cerezo2021variational}. Rather than introducing new ansatz patterns, their work aimed to create a comprehensive catalog of existing quantum circuit ansatzes, facilitating abstraction and reuse in the practical design and implementation of quantum algorithms.

In summary, the research prospects on quantum software architecture might be the following:

\renewcommand{\arealabel}{SA}
\begin{tcolorbox}[colback=LightGray, title=Challenges in Quantum Software Architectures, colframe=DarkGray]
\begin{challenges}
    \item Architectural decisions in quantum software.
    \item Design patterns for hybrid software systems.
    \item Empirical evidence for the application of design patterns.
    \item Evolution of hybrid software architectures.
\end{challenges}
\end{tcolorbox}
\label{ch:sa}

\textbf{Investigating all the factors influencing architectural decisions in quantum software and system design} requires a comprehensive understanding of both the implementation details and the broader technical choices that shape the development of a robust quantum computing system. Exploring these factors should consider a wide range of dimensions, each playing a critical role in guiding the decision-making process. By systematically evaluating these considerations, architects can build systems that not only meet immediate functional requirements but also ensure long-term sustainability and adaptability in an evolving quantum computing landscape.
    
One of the key factors is performance optimization, which involves balancing the limitations and capabilities of quantum hardware, such as gate fidelities, qubit coherence times, and error rates, with the computational goals of the software. Architectural decisions must ensure that quantum algorithms are executed as efficiently as possible, considering the constraints of today’s NISQ devices. This includes minimizing the depth of quantum circuits, optimizing qubit layout to reduce the need for qubit-swapping operations, and employing error correction techniques that don’t overly degrade performance. Performance decisions must also account for how well the quantum system integrates with classical components in hybrid architectures, as seamless interaction between quantum and classical processes is crucial for many near-term applications.
    
Another essential factor is compatibility and interoperability (see also \hyperref[ch:soc]{\textit{Ch-SoC-1}}). As quantum software and systems are developed, architects must ensure that they can function across different quantum hardware platforms and work in tandem with classical systems. This includes making decisions about standardizing communication protocols, programming languages, and intermediate representations like Quantum Intermediate Representation (QIR), ensuring that the quantum architecture remains flexible and capable of adapting to future advancements in quantum hardware. Interoperability also involves considering how the system can support multi-vendor environments, where different components may come from various providers, and how they can collaborate in a seamless manner. This factor is essential for creating scalable, vendor-agnostic solutions that avoid lock-in and provide more flexibility for developers and users.
    
Cost implications are another critical dimension to consider. Quantum computing resources are still expensive, and usage costs can vary based on the number of qubits, the depth of circuits, and the length of time the quantum hardware is in use. Architects must balance these costs with the need for performance, scalability, and experimentation, ensuring that the architecture is not only cost-effective in the short term but also scalable as more affordable quantum technologies become available. Additionally, scalability itself is a significant concern, as quantum systems need to accommodate the expected growth in qubit numbers and system complexity over time. This means planning for architectures that can support larger, more powerful quantum computers in the future, as well as hybrid solutions that evolve with advancements in both quantum and classical computing.

\textbf{Defining the design patterns for building hybrid software systems} that integrate quantum and classical computing is essential for ensuring both scalability and flexibility. At the low-level software component level, the focus should be establishing patterns promoting modularity, reusability, and efficiency. This includes designing compilation units that translate high-level quantum algorithms into executable instructions optimized for various quantum processors. Patterns such as the Factory or Builder patterns can be employed to dynamically create and configure these components, ensuring that quantum circuits and classical routines can be easily reused across different algorithms and applications. Additionally, functions and modules should be encapsulated in a way that allows for the seamless integration of classical and quantum operations while maintaining independence from the underlying hardware (see also \hyperref[ch:soc]{\textit{Ch-SoC-2}}). This ensures that developers can swap or upgrade components without extensive refactoring, promoting code reuse and adaptability in evolving quantum-classical environments.
    
At a high level, architectural design patterns need to be developed to manage the orchestration of services and the execution of workflows that span both quantum and classical systems. This requires patterns that can coordinate the interaction between classical software services and quantum computing tasks \cite{zhao2024towards}. Patterns such as Service-Oriented Architecture (SOA) or Microservices Architecture can be adapted to the quantum domain to manage the distributed nature of hybrid workflows. In these architectures, quantum services act as specialized computational units that can be called upon when needed, and orchestration mechanisms ensure that data flows smoothly between quantum and classical resources. For example, Mediator and Adapter patterns can help manage the communication between QPUs and classical backends, allowing the orchestration layer to interface with multiple quantum platforms without needing to change the core business logic. This pattern-driven approach enables quantum and classical services to operate in tandem, ensuring that workloads are balanced, computational tasks are optimized, and the hybrid system is responsive to varying performance and resource needs.
    
Moreover, workflow management in hybrid systems requires additional patterns that can handle task scheduling, resource allocation, and error management in complex, distributed environments. Patterns like Chain of Responsibility or Observer can be implemented to track the execution of quantum tasks and monitor their status in real-time, allowing the system to react to errors or changes in quantum state coherence. These patterns provide the flexibility needed to manage the dynamic nature of hybrid computations, where quantum resources are expensive and limited, and classical resources are used to support the optimization of quantum workloads. By clearly defining these high-level patterns, developers can ensure that hybrid systems are not only efficient but also modular, scalable, and maintainable, capable of evolving alongside quantum technology advancements. This systematic use of both low-level and high-level design patterns will enable the seamless integration of quantum and classical systems, laying the foundation for scalable, high-performing hybrid computing solutions (cf. Section \ref{sec:services}).

\textbf{Research focused on acquiring substantial empirical evidence for the application of design patterns} in industrial quantum and hybrid computing scenarios is vital for bridging the gap between theoretical advancements and practical implementation. Such research should aim to provide a deeper understanding of how both current and future design patterns and architectural decisions impact the development, deployment, and scalability of quantum systems in real-world settings. The quantum domain, with its rapid evolution and nascent adoption in industry, presents unique challenges that classical software patterns may not fully address. Therefore, empirical studies are essential to validate and adapt these patterns to quantum computing's distinct requirements, such as error rates, qubit coherence, and hybrid quantum-classical interaction.
    
In the immediate term, empirical research should focus on how existing design patterns, adapted from classical computing, function within industrial quantum applications. This involves studying patterns like modularity, reusability, and service orchestration to determine their effectiveness in managing the unique complexities of quantum systems. For example, the research could investigate how well low-level patterns (such as the Factory or Adapter patterns) handle the translation of quantum algorithms across different hardware platforms and whether these patterns enable effective hardware abstraction. Similarly, high-level patterns related to service orchestration and workflow management should be evaluated for their ability to integrate quantum tasks into existing classical infrastructures seamlessly. Real-world case studies involving industries like finance, pharmaceuticals, and logistics, where quantum computing has begun to make an impact, could serve as valuable sources of data for analyzing how these patterns contribute to system efficiency, scalability, and error management.
    
Looking ahead, research should also explore new design patterns and architectural models specifically tailored to quantum computing's future needs. As quantum hardware matures and larger-scale quantum computers become available, new patterns will be required to manage more complex computations and workflows. For example, quantum-native design patterns that account for quantum entanglement, superposition, and error correction should be developed to enhance the effectiveness of hybrid systems. These new patterns might focus on optimizing quantum-classical interactions, reducing quantum circuit depth, or managing qubit connectivity in a way that classical computing patterns cannot address. Research should seek empirical evidence on how these emerging patterns perform in future industrial scenarios, testing them across different quantum hardware platforms, industries, and use cases. Furthermore, this research could identify best practices for adopting these new patterns and inform standardization efforts in the quantum software engineering field, ensuring that quantum solutions are not only powerful but also maintainable and scalable for industrial applications.
    
In addition to examining the direct application of design patterns, research should also consider the broader architectural decisions that guide the development of quantum systems in industry. This includes exploring the impact of architectural choices on performance, cost, and flexibility, as well as the trade-offs between different architectural approaches in hybrid quantum-classical systems. For instance, empirical studies could investigate whether certain architectures are better suited for specific quantum algorithms or whether certain patterns enable more efficient resource allocation and error management in NISQ environments. Understanding how design patterns influence these architectural decisions in practice will help industries make more informed choices about how to integrate quantum computing into their operations. By acquiring empirical evidence on these issues, researchers can provide industries with actionable insights that reduce development risks, streamline quantum adoption, and maximize the potential of quantum computing technologies in real-world scenarios.

\textbf{Investigating the evolution of hybrid software architectures} over time is crucial for understanding the challenges and complexities that arise as quantum and classical systems become more integrated. As these architectures evolve, new patterns, tools, and technologies emerge, which can introduce both improvements and complications. Hybrid quantum-classical systems, in particular, are subject to rapid advancements in quantum hardware and software, which can lead to significant architectural changes. These changes, if not carefully managed, can accumulate technical debt, a situation where shortcuts in design, implementation, or maintenance lead to long-term inefficiencies or scalability issues. Therefore, it is essential to not only track the evolution of these hybrid systems but also develop maintenance strategies that prevent architectural decay and ensure clean, effective, and future-proof designs.
    
One of the key areas to investigate is how the integration of new quantum technologies impacts the long-term health of hybrid architectures. As quantum computing hardware advances, the need to adapt existing architectures to accommodate new qubit types, error correction techniques, or quantum algorithms becomes inevitable. These adaptations may introduce complexity, leading to bloated or inefficient architectures if not managed properly. Researchers should focus on identifying the points at which architectural refactoring is necessary and developing techniques to modernize or restructure hybrid systems without disrupting their core functionality. For example, automated tools could be designed to refactor hybrid architectures by recognizing and replacing outdated components, optimizing circuit depth, or improving the interoperability between quantum and classical services. By tracking how hybrid architectures evolve in response to technological shifts, developers can anticipate points of potential technical debt and address them proactively.
    
In addition to refactoring, hybrid architectures must be continuously maintained to prevent technical debt from accumulating over time. A key strategy for managing technical debt in these systems is the development of automated maintenance solutions. These solutions should be capable of analyzing hybrid architectures for inefficiencies, code duplication, and unused or outdated components, providing developers with actionable insights on how to keep the system clean and effective. For example, maintenance tools could automatically identify quantum circuits that are no longer optimized for the latest hardware or pinpoint classical components that are redundant due to new quantum capabilities. Furthermore, automated testing frameworks that continuously verify the performance, accuracy, and stability of both quantum and classical components could be crucial for preventing the slow buildup of inefficiencies. These tools would help developers mitigate technical debt by providing regular, data-driven feedback on system health, ultimately leading to more sustainable and efficient architectures.
    
Finally, the evolution of hybrid software architectures also demands attention to documentation and version control as a means of ensuring long-term maintainability. As quantum and classical systems evolve together, maintaining accurate and up-to-date documentation becomes increasingly important for managing architectural complexity. Documentation should cover both the high-level architecture of the system and the details of how quantum and classical components interact, ensuring that future developers can easily understand and extend the system without introducing additional complexity. Version control systems must also be adapted to manage the distinct versions of quantum software, classical software, and their integration points, ensuring that any changes to the system are well-tracked and reversible if needed. Establishing best practices for maintaining clear and consistent documentation, along with robust versioning strategies, will be essential to keeping hybrid architectures maintainable as they evolve over time, thereby reducing technical debt and enhancing long-term system effectiveness.

%%%%%%%%%%%%%%%%%%%%%%%%%%%%%%%%%%%%%%%%%%%%%%%%%%
\subsection{Software Development Processes}
\label{sec:processes}

Akbar \textit{et al.} \cite{Akbar2023} categorize various QSE challenges. Specifically, two challenges suggest that specific quantum/hybrid software development processes are needed: the 14th challenge: ``\textit{Integration with classical computing}'', and the 22nd challenge: ``\textit{Project management issues}''. With regard to the integration of classical quantum software, quantum software cannot be developed and operated in isolation. ``\textit{A key challenge is to fully integrate these types of the system into a unified classical and quantum software development lifecycle}'' \cite{SEI2021}. Thus, there are some issues, for example, the interpretation of the results by the classical counterpart, since quantum computations are stochastic \cite{Haghparast23}. In addition, new architectural paradigms and design patterns will be necessary to develop quantum software effectively \cite{Khan23}. Regarding project management, some issues, such as risk management, are specifically framed in the development of hybrid software systems. For example, the limited availability of real quantum hardware, which leads to testing quantum software in simulators, sometimes leads to different results when the quantum software is executed in the production environment. Other risks with a high impact on the success of quantum software development projects are the lack of standardized tools and frameworks, or scalability \cite{Akbar2024}.

There is some preliminary research on the holistic quantum software development lifecycle. Thus, Zhao \cite{Zhao2020} proposed a slight adaptation of the waterfall model to support the design and construction of quantum software systems, covering phases such as requirements analysis, design, implementation, testing, and maintenance. In a similar approach based on the waterfall model, Dey \textit{et al.} \cite{dey2020qdlc} proposed some alternative stages.

Since these models can inherit all the weaknesses of the classical waterfall life cycle, other proposals are based on iterative models. Weder \textit{et al.} \cite{Weder20} proposed an iterative model based on ``quantum data provenance'', which serves in different phases of the life cycle. In addition, Pérez-Castillo \textit{et al.} \cite{ICSM2024} propose an adaptation of the Incremental Commitment Spiral Model (ICSM), which is also an iterative model that suggests risk management during software development. 

Other proposals examine how agile good practices can be integrated into quantum software development. In this context, proposals have emerged that address the integration of the DevOps paradigm within the domain of quantum software development \cite{Stirbu23, GheorghePop2020QuantumDT, romero2023enabling}. In a more analytical way, an interview-based study \cite{Akbar2022} shows the most important challenges in agile-quantum software development, such as sustainable scaling and the need for mature tool ecosystems and standard agile specifications. The significance of these challenges is likely to be paramount in the coming years. 

The open research topics for the next years in this domain might be the following:

\renewcommand{\arealabel}{DP}
\begin{tcolorbox}[colback=LightGray, title=Challenges in Quantum Software Development Processes, colframe=DarkGray]
\begin{challenges}
    \item Iterative development of hybrid software.
    \item Risk management.
    \item Project management.
\end{challenges}
\end{tcolorbox}
\label{ch:dp}

\textbf{Managing iterative development of hybrid software systems} through the whole lifecycle. Iterative development models, which have proven effective in managing complexity and enhancing flexibility in classical software projects, must be adapted to hybrid software systems development. This includes models that can accommodate rapid changes in quantum technology and provide frameworks for the co-development of classical and quantum components (see also \hyperref[ch:sa]{\textit{Ch-SA-4}}). This, in turn, will introduce unique challenges, such as the need for quantum-specific design patterns, integration testing frameworks, and development tools that can handle quantum-classical software systems. 

\par A survey conducted by Jiménez-Navajas \textit{et al.} \cite{JimenezNavajas2024} shows that there is a certain dissatisfaction with specific tools, indicating a gap between the available tools and the developers’ needs. In particular, that survey indicates that current modeling tools are insufficient for quantum and hybrid quantum-classical software. While classical software has more robust modeling support, there is a significant need for better tools to model quantum and hybrid software systems effectively. This is aligned with MDE challenges presented in Section \ref{sec:MDE}.

\textbf{Risk management} within the specialized realm of quantum software, coupled with sustainable scalability strategies, is essential for developing progressively larger and more intricate hybrid software systems. Thus, adaptive risk management strategies, which can evolve with the quantum computing landscape, will be necessary. This might involve dynamic resource allocation models, contingency planning for quantum hardware failures, and methodologies for evaluating the reliability of quantum algorithms. 
    
\par In addition to the risks involved in quantum computing \textit{per se} \cite{WEF2022}, there are some specific risks for quantum software development \cite{Hevia2024}. Table \ref{tab:risksTable} summarizes most of the risks mentioned above, pointing out some approaches that can be used to mitigate them. As we can see, good practices in software and systems engineering (MDE, SOA, ADM, metrics, etc.) can be used to mitigate some of these risks.

    \begin{table}[h!]
    \caption{Main quantum software risks and mitigation approaches adapted from \cite{Hevia2024}}.
    \label{tab:risksTable}
    \centering
    \begin{tabularx}{\textwidth}{|X|X|}
    \hline
    \multicolumn{1}{|c|}{\textbf{Risk}} & \multicolumn{1}{c|}{\textbf{Mitigation approach}} \\ \hline
    Lack of preparation & Training, automation (AQSE) \\ \hline
    Difficulty in use & User-friendly, Graphical User Interfaces (GUIs) \\ \hline
    Algorithm complexity & Training, algorithm libraries \\ \hline
    Variety of types of quantum computers & Technology agnosticism \\ \hline
    Changes in the same quantum system & Full portability of   quantum software \\ \hline
    Diversity of programming languages & MDE, low code techniques \\ \hline
    Integration of   classical and quantum IT & Service Oriented   Architecture (SOA), standardized APIs \\ \hline
    Hybrid information systems construction & New hybrid software   life cycles \\ \hline
    Migration of classical software & Quantum software   modernization (ADM, Architecture-Driven Modernization) \\ \hline
    Poor quality of quantum software & New testing techniques/New software quality characteristics and metrics \\ \hline
    Problems in execution and results/Environment management & An integrated environment for design and execution \\ \hline
    Lack of community & Create quantum software networks and interest groups \\ \hline
    \end{tabularx}
    \end{table}

\par One of the most crucial risk mitigation strategies is adopting an agnostic architecture, which helps safeguard the investments that users have made in developing quantum software assets. 
    
\par We are experiencing a continuous and often rapid evolution in the disruptive field of quantum computing. Given the current state of development, making precise predictions regarding the widespread adoption and deployment of general-purpose quantum computing remains difficult \cite{Deshpande2022}. As a result, it is even more critical in quantum computing than in other areas of information technology to conduct thorough risk assessments. The responsibility will fall to the ``\textit{Quantum Business Strategist}'' \cite{Lenahan2021a} to develop a risk management strategy grounded in an integrated approach, allowing organizations to understand quantum risks comprehensively.

\textbf{Project management}, particularly in its focus on the operational aspects of quantum software, requires the integration of DevOps or similar agile paradigms into the software development lifecycle. Hence, future research lines could focus on developing or adapting agile toolchains, including version control systems, continuous integration/ continuous deployment (CI/CD) pipelines, and quantum-aware agile project management tools. This research line could investigate how quantum computing as a service can be integrated into software development projects, enabling agile teams to leverage quantum computing resources efficiently.

\par In addition, another critical point during project management is the impact of automated code generation tools tailored to quantum software engineering. The use of LLMs to assist in software development has shown promising results in classical software, and it is expected that similar techniques could be applied to quantum software, as demonstrated in some preliminary works \cite{dupuis2024, guo2024repairing}. LLMs could assist developers in writing, refactoring, and optimizing quantum code, potentially lowering the barrier to entry for quantum software development and accelerating the development process (see Section \ref{sec:ai}). However, challenges in using large language models (LLMs) for software development include bias and inaccuracy in generated code, security risks from suggesting vulnerable patterns, and limitations in contextual understanding, which can lead to integration issues within complex projects. Additionally, there are ethical and legal concerns around intellectual property, especially when LLMs may reproduce licensed code. Developers also face a learning curve in effectively using these tools, and integration into existing development workflows, such as CI/CD pipelines, remains a challenge. Finally, trust in AI-generated code requires careful human oversight to ensure reliability.

%%%%%%%%%%%%%%%%%%%%%%%%%%%%%%%%%%%%%%%%%%%%%%%%%%
\subsection{Artificial Intelligence}
\label{sec:ai}

As in almost every software engineering domain, QSE has been increasingly influenced by the recent advances in artificial intelligence (AI). The intersection between AI and QSE is a promising and rapidly growing area of research. On the one hand, applying AI techniques to enhance the processes and methodologies within QSE is starting to yield significant benefits. AI can assist in optimizing quantum software development and automating tasks like quantum circuit optimization, error correction, and hybrid system orchestration. On the other hand, quantum computing itself holds immense potential to revolutionize AI, particularly by developing quantum algorithms designed to enhance machine learning and deep learning processes, leading to the exploration of Quantum AI systems.

In terms of using AI for Quantum Software Engineering, several research teams are actively exploring how machine learning models can help design, test, and optimize quantum algorithms. For instance, AI can automatically suggest optimized quantum circuits or identify more efficient qubit allocations, thereby improving the overall performance of quantum computations. AI-driven tools can also aid in error mitigation by predicting and compensating for quantum noise, which is especially important in NISQ devices. Additionally, AI-enhanced simulators could provide more accurate predictions of quantum system behaviors, helping developers better understand the limitations of their algorithms and hardware and adjust accordingly. AI can also be integrated into quantum programming environments to assist with debugging, verification, and testing of quantum circuits, areas that are notoriously complex in quantum software.

Works such as the one proposed by Salm \textit{et al.} \cite{salm2022optimizing} propose using machine learning approaches to optimize the selection of compiled quantum circuits for the available quantum computers. This work is further developed in \cite{salm2023select}, where several machine learning algorithms are used to improve such selection. A similar approach was also followed by Garcia-Alonso \textit{et al.} \cite{garcia2021quantum}, where a predictive model is used to determine the minimal waiting time for executing a quantum circuit.

Several works have also been developed in the domain of Variational Quantum Algorithms that try to leverage the benefits of AI. For example, Furutanpey \textit{et al.} \cite{furutanpey2023architectural} propose using Quantum Neural Networks to help system designers distribute hybrid quantum applications through the computing continuum. Similarly, in \cite{truger2022selection}, the authors address the problem of hyperparameter selection in warm-starting quantum optimization problems.

In addition, neural networks have also been used for the automated synthesis of quantum circuits \cite{murakami2022automated}. This work proposes AutoQC, a tool for synthesizing quantum circuits using the neural network from input and output pairs. In the same line, Dupuis \textit{et al.} \cite{dupuis2024} focus on training code LLMs to generate quantum computing code, addressing challenges such as limited datasets and evolving technology, and demonstrating superior performance in tasks using the Qiskit library. From a different point of view, Quantum Machine Learning approaches have been empirically evaluated from a Software Engineering perspective in order to study the bugs present in such approaches. Also, Zhao \textit{et al.} \cite{zhao2023empirical} try to ensure the correctness and robustness of Quantum Machine Learning approaches by inspecting almost 400 real-world bugs collected from more than 20 repositories of popular QML frameworks. And, of course, the nowadays ubiquitous ChatGPT has also been analyzed in terms of its ability to repair quantum programs \cite{guo2024repairing}.

Conversely, Quantum Software Engineering for AI is another burgeoning field, with research teams investigating how quantum computing can be leveraged to improve AI algorithms. Quantum computing’s ability to process vast amounts of data in parallel through superposition and entanglement has the potential to enhance machine learning algorithms significantly. Quantum versions of AI algorithms, such as quantum neural networks and quantum support vector machines, are expected to be more efficient than their classical counterparts in processing large datasets and solving complex optimization problems (although loading classical data presents a challenge \cite{miranskyy2024comparing}). The unique characteristics of quantum computing, such as quantum entanglement and superposition, may enable more powerful AI models that can learn faster and handle more complex patterns than classical AI systems. As a result, QSE is playing an essential role in developing the tools, methodologies, and frameworks necessary to design and implement quantum-enhanced AI systems, making it a critical enabler of the next generation of AI technology.

Overall, the synergy between AI and Quantum Software Engineering opens up exciting possibilities for innovation in both fields. By combining AI’s capacity to learn, optimize, and automate with quantum computing’s unparalleled computational power, researchers are beginning to push the boundaries of what is achievable in both quantum software development and artificial intelligence. This intersection promises to drive progress in solving some of the most complex problems in computing, from optimizing quantum hardware to creating more intelligent and efficient AI models. As research continues to expand in this direction, the integration of AI and QSE will likely become a cornerstone of technological advancements in both quantum computing and artificial intelligence, unlocking new applications and accelerating progress in these cutting-edge domains.

These are just some examples of the initial impact of the relationship between Quantum Software engineering and AI on the research community. Nevertheless, additional challenges need to be addressed in this domain. Specifically:

\renewcommand{\arealabel}{AI}
\begin{tcolorbox}[colback=LightGray, title=Challenges in Quantum Artificial Intelligence, colframe=DarkGray]
\begin{challenges}
    \item Quantum Circuit Optimization.
    \item Developing Hybrid AI-Quantum Workflows.
    \item Error Mitigation and Correction.
    \item Scalability of AI-Assisted Quantum Software Development.
\end{challenges}
\end{tcolorbox}
\label{ch:ai}

\textbf{Quantum Circuit Optimization with AI.} Quantum circuits are often highly complex and require optimization to run efficiently on current quantum hardware, with limitations such as qubit coherence time and gate fidelities. Although AI techniques, such as reinforcement learning and evolutionary algorithms, have shown potential in optimizing classical software, applying these techniques to quantum circuits presents significant challenges. AI-driven optimizations must account for quantum-specific issues like gate depth, noise resilience, and the limited number of qubits. The research community must focus on developing AI models that can handle quantum-specific characteristics while ensuring the optimized circuits are scalable and executable on real quantum devices. Moreover, techniques that reduce the time it takes for AI systems to learn and improve quantum circuit designs need to be explored, making the process both time-efficient and cost-efficient.

\textbf{Developing Hybrid AI-Quantum Workflows.} Quantum computing's current state requires hybrid workflows involving both classical and quantum components. The challenge lies in integrating AI techniques within these hybrid quantum-classical systems. Designing efficient workflows where AI can dynamically decide which parts of a computation should be offloaded to quantum systems and which should remain classical is not trivial. AI algorithms capable of predicting the best partitioning of tasks between quantum and classical systems based on available resources and computational goals are needed. Additionally, these workflows must optimize communication between quantum and classical systems, balancing latency and data transfer speeds. The research community must investigate how AI can be effectively integrated into quantum software frameworks to create adaptive, efficient, and seamless hybrid systems.

\textbf{AI for Quantum Error Mitigation and Correction.} Quantum systems are inherently noisy, and error correction is one of the most significant challenges facing quantum computing. While AI has successfully automated certain aspects of error detection and correction in classical computing, the unique nature of quantum errors, such as decoherence and gate errors, requires new approaches. AI models could potentially be trained to predict quantum system errors and suggest real-time corrections. However, building AI systems that can learn and adapt to the specific noise models of quantum processors and account for the probabilistic nature of quantum measurements presents a significant research challenge. The scientific community must explore how AI techniques can be tailored to quantum error mitigation strategies while ensuring they work efficiently within the constraints of current quantum hardware.

\textbf{Scalability of AI-Assisted Quantum Software Development.} As quantum computing systems scale, so must the software and algorithms that control them. AI has the potential to accelerate quantum software development, but ensuring that AI-driven development processes scale effectively is a major challenge. Quantum software engineering currently lacks mature development environments, automated tools, and debugging support at the scale needed for widespread industry use. Developing AI systems that assist in code generation, automated testing, and debugging for quantum systems is essential, but these systems must scale as quantum computers grow in power and complexity. The research community should focus on building robust AI-assisted frameworks that can handle larger quantum systems, addressing issues such as increased circuit complexity, debugging for hybrid systems, and the integration of AI across multiple layers of quantum software.

%%%%%%%%%%%%%%%%%%%%%%%%%%%%%%%%%%%%%%%%%%%%%%%%%%
\subsection{Summary of the different key challenges of quantum software engineering}
\label{sec:summary}

A summary table of the different key challenges of quantum software engineering seen throughout this section is provided below:

\begin{tcolorbox}[colback=LightGray, title=Summary, colframe=DarkGray]

    \begin{itemize}
        \renewcommand{\arealabel}{ST}
        \item \underline{\textbf{Challenges in Quantum Software Testing}} (\ref{sec:testing})
        \begin{challenges}
            \item Efficient test oracles.
            \item Test scalability.
            \item From simulators to real quantum computers.
            \item (Quantum) Artificial Intelligence (AI) and (Quantum) Software Testing.
        \end{challenges}

        \renewcommand{\arealabel}{SoC}
        \item \underline{\textbf{Challenges in Quantum Service-Oriented Computing}} (\ref{sec:services})
        \begin{challenges}
            \item Interoperability.
            \item Platform independence.
            \item Demand and Capacity Management.
            \item Workforce training.
        \end{challenges}

        \renewcommand{\arealabel}{MDE}
        \item \underline{\textbf{Challenges in Quantum Model-Driven Engineering}} (\ref{sec:MDE})
        \begin{challenges}
            \item Modelling quantum-specific constructs.
            \item Development of high-level design methodologies.
            \item Scalable quantum software maintenance and evolution.
            \item Intelligent code generation and orchestration.
        \end{challenges}

        \renewcommand{\arealabel}{PP}
        \item \underline{\textbf{Challenges in Quantum Programming Paradigms}} (\ref{sec:paradigms})
        \begin{challenges}
            \item Complexity of circuits.
            \item Composable and reusable quantum software.
            \item Abstractions for quantum software.
        \end{challenges}

        \renewcommand{\arealabel}{SA}
        \item \underline{\textbf{Challenges in Quantum Software Architectures}} (\ref{sec:architecture})
        \begin{challenges}
            \item Architectural decisions in quantum software.
            \item Design patterns for hybrid software systems.
            \item Empirical evidence for the application of design patterns.
            \item Evolution of hybrid software architectures.
        \end{challenges}

        \renewcommand{\arealabel}{DP}
        \item \underline{\textbf{Challenges in Quantum Software Development Processes}} (\ref{sec:processes})
        \begin{challenges}
            \item Iterative development of hybrid software.
            \item Risk management.
            \item Project management.
        \end{challenges}

        \renewcommand{\arealabel}{AI}
        \item \underline{\textbf{Challenges in Quantum Artificial Intelligence}} (\ref{sec:ai})
                \begin{challenges}
            \item Quantum Circuit Optimization.
            \item Developing Hybrid AI-Quantum Workflows.
            \item Error Mitigation and Correction.
            \item Scalability of AI-Assisted Quantum Software Development.
        \end{challenges}
        
    \end{itemize}
\end{tcolorbox}

\section{Software engineering key areas and challenges regarding quantum computing}
\label{sec:keyAreas}

The landscape of software engineering has undergone a profound transformation in recent years, driven by the rapid evolution of new technologies, methodologies, and computing paradigms. As a result, there is a growing need for a new roadmap reflecting the future research direction in this dynamic field. A key example of this shift is the emergence of quantum software engineering, which presents unique challenges and demands fundamental changes in the core principles of software engineering \cite{Zhao2020}. Quantum computing, as we have already seen, introduces novel concepts such as superposition, entanglement, and probabilistic behavior, all requiring rethinking traditional software engineering paradigms to accommodate the distinct characteristics of quantum systems. The impact of these changes is expected to reshape not only quantum software engineering but the broader landscape of software development as a whole.

Already, researchers and thought leaders are actively considering what the short-term to medium-term future of software engineering will look like, both in general and specifically within the realm of QSE \cite{Murillo2024}. A notable event in this context is the \textit{2030 Software Engineering Roadmap workshop} that was co-located with \textit{ACM SIGSOFT FSE Foundations of Software Engineering} on July 15th and 16th, 2024 in \textit{Porto de Galinhas, Brazil}\footnote{2030 Software Engineering Roadmap workshop. \url{https://conf.researchr.org/home/2030-se}}. This workshop brought together researchers worldwide to discuss the key challenges facing software engineering in the current decade. During the event, participants explored recent shifts in software engineering practices, shared their vision of the field’s future evolution, and collaborated on designing a roadmap to guide the research community toward addressing these emerging challenges.

Drawing on the insights from the workshop \cite{Pezze2024}, six key areas were identified as the most pressing challenges for software engineering research going forward. These areas reflect both the broader concerns within traditional software engineering and the specialized issues that arise in the context of quantum computing. In this section, we will outline these six key areas in detail, focusing on how they intersect with the field of QSE and how advancements in quantum computing may shape the future of software engineering as a whole.

\textbf{Artificial Intelligence for Software Engineering (AI4SE).} The recent breakthroughs in machine learning, generative AI, and autonomous systems represent the most profound transformation in software engineering research and practice since the advent of the Internet in the latter half of the 20th century \cite{Terragni2024}. The software engineering community has never experienced such a rapid and dominant rise in new research directions, with topics like machine learning in software engineering and the challenges of engineering AI-driven systems becoming central themes in leading conferences and journals \cite{Jackson2024}.
    
The relationship between Artificial Intelligence and Quantum Computing offers immense potential to transform both fields, particularly in the context of software engineering (see also \hyperref[ch:ai]{\textit{Ch-AI}}). As AI becomes more integrated into software development practices, it can significantly impact the evolution of quantum computing by providing advanced tools for algorithm design, circuit optimization, and error management. AI can help automate the design of quantum circuits, optimize qubit usage, and even predict and correct quantum errors in real-time, all of which are essential for improving the efficiency and reliability of quantum computations. This synergy is especially important in hybrid quantum-classical systems, where AI can manage the distribution of tasks between classical and quantum processors, ensuring the efficient orchestration of computational resources. Using AI to enhance the development process, quantum computing systems can become more scalable and accessible to a broader range of developers and industries.
    
On the other hand, quantum computing offers the potential to revolutionize AI itself, particularly in the domain of AI4SE. Quantum computing's ability to process large amounts of data simultaneously through quantum superposition and entanglement could dramatically speed up key machine learning algorithms, allowing AI models to be trained faster and more efficiently. This would enable AI systems to handle larger datasets and solve more complex optimization problems, making AI-driven tasks more powerful and accurate. Furthermore, quantum computing could allow AI models to better understand and manage complex software systems by analyzing multiple variables and configurations at once, leading to deeper insights into system behavior and improved software reliability.

\textbf{Software Engineering by and for Humans (SE4H).} Machine learning, Generative AI, and autonomous systems are reshaping the landscape of software engineering, fundamentally altering the traditional concept of software artifacts \cite{Ciniselli2024}. These emerging technologies introduce complex ethical, fairness, and technical challenges that software engineers must now navigate. Humans are no longer just users of software systems, they have become integral components of expansive cyber-physical ecosystems. As a result, the scope of software engineering research must expand beyond the narrow view of human users and embrace a broader perspective that considers humans as essential elements within these interconnected systems \cite{Wang2024}.

SE4H strongly emphasizes creating software systems that are ethical, user-friendly, and aligned with human values. This approach can significantly influence the development of Quantum Computing by shaping how quantum systems are designed, implemented, and deployed. Quantum computing has the potential to solve complex problems that classical computing cannot handle efficiently, but the power and complexity of these systems also raise significant concerns related to transparency, accessibility, and usability. SE4H can drive the development of quantum systems that prioritize human-centric design, ensuring that quantum applications are not just powerful but also understandable, ethical, and responsive to user needs. For instance, SE4H principles can guide the creation of quantum software interfaces that simplify interactions between users and quantum machines, making quantum computing more approachable for non-experts while maintaining the necessary levels of control and oversight.
    
Conversely, quantum computing can profoundly influence software engineering for humans by introducing new capabilities for solving problems that directly impact human-centered applications. Quantum computing could revolutionize areas like healthcare, environmental modeling, and cryptography, enabling more efficient solutions to problems that have been computationally intractable using classical methods. These breakthroughs can enhance software systems designed for human use by providing faster, more accurate, and scalable solutions. However, as quantum computing introduces novel algorithms and computing paradigms, SE4H will need to evolve to ensure that quantum software is developed with consideration for human ethics, usability, and societal impact. Quantum computing also requires new models for how humans interact with software, given the complexity of quantum mechanics, making SE4H crucial in ensuring that these systems are accessible and understandable to diverse user groups.

\textbf{Automatic Programming.} Machine learning, particularly through deep neural networks and large language models, represents the most significant amplification of human productivity in software engineering since its early days. These technologies are pushing the boundaries of what is possible by enabling new frontiers in automatic programming and transforming how software is written, tested, and maintained. They are also reshaping the landscape of quality and security, raising critical questions about how we ensure robust, secure, and reliable systems in the face of AI-generated code. These advancements also introduce new societal and legal issues, such as accountability and the ethical implications of automation in software development.

Automatic programming, using AI to generate code with minimal human input, has the potential to influence quantum computing development significantly. Quantum programming is inherently more complex than classical programming due to the nature of quantum mechanics, involving concepts like superposition, entanglement, and probabilistic outcomes. These factors make quantum software development highly specialized and challenging. Automatic programming can help bridge this gap by generating quantum code from high-level descriptions, allowing developers who may not be experts in quantum computing to leverage the power of quantum computing. AI-driven tools can automate the creation and optimization of quantum circuits, reduce human error, and speed up the development process, making quantum computing more accessible to a broader range of developers and industries (see also \hyperref[ch:ai]{\textit{Ch-AI}}).
    
On the other hand, quantum computing can influence automatic programming by enhancing the performance of AI algorithms used to generate code. Quantum computers have the potential to dramatically improve the efficiency of machine learning models that underlie automatic programming systems. Quantum algorithms could be used to optimize the search spaces involved in code generation, enabling faster and more efficient solutions to programming challenges. Additionally, quantum computing could allow automatic programming systems to tackle more complex problems, such as large-scale software optimization or real-time bug fixing, that are currently too resource-intensive for classical computing.

\textbf{Software Security.} The widespread integration of machine learning in software quality and security brings new societal and legal considerations \cite{Molina2024}. As software systems become increasingly complex and large-scale, novel security challenges arise, underscoring the need for advanced methods in secure software engineering and cybersecurity \cite{Patnaik2024}.

Software security plays a vital role in the development of quantum computing, as the rise of quantum technologies introduces new vulnerabilities and security risks. With quantum computers having the potential to break widely used encryption algorithms, particularly those based on classical public-key cryptography, securing quantum software becomes an urgent priority. As quantum systems evolve, the integration of security measures, such as quantum-resistant cryptographic algorithms \cite{BarzenAndLeymann2024,zhang2021quantum,zhang2023making} and secure quantum communication protocols, will be crucial in protecting data and applications from breaches. Software security methodologies, including secure coding practices, vulnerability detection, and real-time threat monitoring, will need to adapt to the unique characteristics of quantum environments (see also \hyperref[ch:st]{\textit{Ch-ST}}), ensuring that quantum applications and their interactions with classical systems are secure from new types of cyber threats.
    
On the other hand, quantum computing can transform software security by enabling more advanced cryptographic techniques and faster detection of vulnerabilities. Quantum algorithms, such as Shor’s algorithm, have the ability to factor large integers exponentially faster than classical algorithms, which poses a threat to current encryption methods. However, quantum cryptography, particularly quantum key distribution, offers an unprecedented level of security by leveraging the principles of quantum mechanics to detect eavesdropping and ensure secure communication channels. Additionally, quantum computing could enhance the speed and accuracy of security tools, allowing for the rapid identification of software vulnerabilities and the development of more robust defense mechanisms against emerging cyber threats.

\textbf{Validation and Verification (V\&V).} Machine learning and AI are transforming the landscape of validation and verification in software engineering. The adaptive and evolving nature of AI-driven systems changes the traditional concepts of test input and oracles, as these systems can continuously learn and modify their behavior. Simultaneously, machine learning and generative AI open up new possibilities for automating the testing process, offering innovative approaches to test generation and analysis. There is a pressing need for a new conceptual framework that addresses the unique challenges of testing and analyzing AI-driven software systems, as well as further research into how ML and generative AI can be harnessed to improve testing and analysis.

V\&V are critical aspects of software development, ensuring that systems perform as intended and meet specified requirements. In quantum computing, V\&V methodologies are essential for managing the inherent complexity and probabilistic nature of quantum algorithms. As quantum programs deal with qubits, superposition, and entanglement, traditional V\&V methods must evolve to account for the unique characteristics of quantum systems. This includes developing tools for verifying the correctness of quantum circuits, identifying quantum-specific bugs, and ensuring that quantum systems behave reliably across different execution environments (see also \hyperref[ch:st]{\textit{Ch-ST}}). Enhanced V\&V processes will play a crucial role in building trust in quantum applications, particularly as they become integrated into high-stakes domains like cryptography, pharmaceuticals, and optimization.
    
Conversely, quantum computing has the potential to revolutionize validation and verification by providing unprecedented computational power to handle complex verification tasks. Quantum computers can process vast amounts of data in parallel, potentially allowing them to verify intricate software systems far more efficiently than classical methods. For example, quantum algorithms could dramatically accelerate the analysis of software models, enabling faster identification of bugs, inconsistencies, or security vulnerabilities in both quantum and classical software. This could lead to the development of quantum-enhanced verification tools capable of validating complex configurations and simulations that are computationally intractable for classical systems. In this way, quantum computing can push the boundaries of what is possible in V\&V, enabling deeper and more comprehensive analysis of large-scale software systems.

\textbf{Sustainable Software Engineering.} The concept of sustainable development now extends beyond traditional environmental concerns and encompasses software systems operating within cyber-physical spaces. Achieving sustainability in these systems requires innovative approaches to design, development, deployment, and maintenance that prioritize reducing ecological impact, improving resource efficiency, and promoting social responsibility.

Sustainable Software Engineering can influence the development of quantum computing by encouraging the design of quantum systems that minimize energy consumption and optimize resource usage. As quantum computing evolves, it is essential to address its potential environmental impact, particularly in the context of large-scale quantum processors that may require significant amounts of energy for cooling and operation. By applying principles of sustainability, researchers can develop quantum software and hardware that prioritize energy efficiency and minimize the ecological footprint. This could involve optimizing quantum algorithms to run on fewer qubits or for shorter durations, reducing the overall power consumption of quantum systems and contributing to more environmentally responsible computing practices (see also \hyperref[ch:pp]{\textit{Ch-PP}} and \hyperref[ch:dp]{\textit{Ch-DP}}).
    
Conversely, quantum computing has the potential to reshape Sustainable Software Engineering by providing solutions to problems that are currently too complex or resource-intensive for classical computing. Quantum computing’s ability to solve optimization problems more efficiently can lead to breakthroughs in areas such as energy distribution, climate modeling, and resource management. In this way, quantum computing could directly support the goals of sustainable development by providing more efficient solutions to critical sustainability challenges.

%%%%%%%%%%%%%%%%%%%%%%%%%%%%%%%%%%%%%%%%%%%%%%%%%%
%%%%%%%%%%%%%%%%%%%%%%%%%%%%%%%%%%%%%%%%%%%%%%%%%%
\section{Conclusion and challenges}
\label{sec:conclusion}

% \note{New conclusions and challenges.}

In this work, a group of active researchers is currently addressing the challenges of Quantum Software Engineering and analyzing the most recent advances in this field. This analysis reveals some certainties, which can be summarized as follows. First, no matter the discipline one focuses on, quantum software development requires new techniques compared to classical software development. These techniques have already begun to form the body of QSE. Second, quantum computers are reaching a development stage attracting industry attention. Therefore, empirical software engineering methods and techniques need to be revisited to meet industry expectations regarding quantum computing, making the development of QSE a priority.

To achieve this development, QSE's most noteworthy challenges have been identified. Next, it is necessary to identify synergies and dependencies between them further. This could help researchers in the QSE domain to focus on the most needed aspects. By successfully addressing these challenges, QSE will be able to support the development of hybrid software systems beyond the NISQ era. The roadmap does not end here: quantum technology will continue to evolve, posing new challenges and opportunities for the QSE community.

%%%%%%%%%%%%%%%%%%%%%%%%%%%%%%%%%%%%%%%%%%%%%%%%%%
%%%%%%%%%%%%%%%%%%%%%%%%%%%%%%%%%%%%%%%%%%%%%%%%%%
\section*{Acknowledgments}

S. Ali is supported by the Qu-Test project (Project \#299827) funded by the Research Council of Norway and Simula's internal strategic project on quantum software engineering. P. Arcaini is supported by Engineerable AI Techniques for Practical Applications of High-Quality Machine Learning-based Systems Project (Grant Number JPMJMI20B8), JST-Mirai. A. Brogi was partly supported by project UNIPI PRA 2022 64 “OSMWARE”, funded by the University of Pisa, Italy. R. Pérez-Castillo, I. García Rodríguez de Guzmán and M. Piattini are supported by the projects Q-SERV (PID2021-124054OB-C32), SMOOTH (PID2022-137944NB-I00) and QU-ASAP (PDC2022-133051-I00) funded by MICIU/AEI/ 10.13039/501100011033 / PRTR, EU. A. Miranskyy is supported by the Natural Sciences and Engineering Research Council of Canada (Grant Number RGPIN-2022-03886). Juan M. Murillo, Jose Garcia-Alonso and Enrique Moguel are supported by the projects Q-SERV (PID2021-124054OB-C31). RuralServ (TED2021-130913B-I00), PDC2022-133465-I00, and RCIS (RED2022-134148-T) funded by the Ministry of Science and Innovation. They are also supported by the project ``TECH4E -Tech4effiency EDlH (101083667)'' supported by the European Commission through the Digital Europe Program. Antonio Ruiz-Cortés is supported by the Perseo Project (PID2021-126227NB-C21) and RCIS (RED2022-134148-T) funded by the Ministry of Science and Innovation. Jianjun Zhao is supported by JSPS KAKENHI Grant No. JP23H03372, and No. JP24K14908. T. Yue is supported by the State Key Laboratory of Complex \& Critical Software Environment (SKLCCSE, grant No. CCSE-2024ZX-01) and the Fundamental Research Funds for the Central Universities.

%\begin{acks}
%To Robert, for the bagels and explaining CMYK and color spaces.
%\end{acks}

\bibliographystyle{ACM-Reference-Format}
\bibliography{references}

%\appendix

%\section{Appendix A}

\end{document}